\documentclass[aps,prb,twocolumn,showpacs,superscriptaddress]{revtex4-1}
\usepackage{graphicx}
\usepackage{epsfig}
\usepackage{latexsym}
\usepackage[dvips]{color}

\begin{document}

\title{Quadrupolar Singlet Ground State of Praseodymium in a Modulated Pyrochlore}

\author{J. van Duijn}
\affiliation{Department of Physics and Astronomy, John Hopkins University, Baltimore, Maryland 221218}
\affiliation{ISIS Facility, Rutherford Appleton Laboratory, Chilton, Didcot, OX11 0QX, U.K.}
\affiliation{Instituto de Investigaci\'{o}n en Energ\'{i}as Renovables, Departamento de F\'{i}sica Aplicada, Universidad de Castilla-La Mancha, Albacete, 02006, Spain}
\author{K. H. Kim}
\affiliation{Rutgers Center for Emergent Materials and Department of Physics and Astronomy, Rutgers University, Piscataway, New Jersey 08854}
\author{N. Hur}
\affiliation{Rutgers Center for Emergent Materials and Department of Physics and Astronomy, Rutgers University, Piscataway, New Jersey 08854}
\author{D. T. Adroja}
\affiliation{ISIS Facility, Rutherford Appleton Laboratory, Chilton, Didcot, OX11 0QX, U.K.}
\author{F. Bridges}
\affiliation{Department of Physics, University of California, Santa Cruz, California 95064}
\author{A. Daoud-Aladine}
\affiliation{ISIS Facility, Rutherford Appleton Laboratory, Chilton, Didcot, OX11 0QX, U.K.}
\author{F. Fernandez-Alonso}
\affiliation{ISIS Facility, Rutherford Appleton Laboratory, Chilton, Didcot, OX11 0QX, U.K.}
\affiliation{Department of Physics and Astronomy, University College London, Gower Street, London, WC1E 6BT, U.K}
\author{R. Ruiz-Bustos}
\affiliation{Departamento de Mec\'{a}nica, Universidad de C\'{o}rdoba, C\'{o}ordoba, 14071, Spain}
\author{Jiajia Wen}
\affiliation{Institute for Quantum Matter and Department of Physics and Astronomy, Johns Hopkins University, Baltimore, Maryland 21218}
\author{ V. Kearney}
\affiliation{Department of Physics, University of California, Santa Cruz, California 95064}
\author{Q. Z. Huang}
\affiliation{NIST Centre for Neutron Research, National Institute of Standards and Technology, Gaithersburg, Maryland 20899}
\author{S.-W. Cheong}
\affiliation{Rutgers Center for Emergent Materials and Department of Physics and Astronomy, Rutgers University, Piscataway, New Jersey 08854}
\author{S. Nakatsuji}
\affiliation{Institute for Solid State Physics, University of Tokyo, Kashiwa, Chiba 277-8581, Japan}
\author{C. Broholm}
\affiliation{Institute for Quantum Matter and Department of Physics and Astronomy, Johns Hopkins University, Baltimore, Maryland 21218}
\affiliation{NIST Centre for Neutron Research, National Institute of Standards and Technology, Gaithersburg, Maryland 20899}
\author{T. G. Perring}
\affiliation{ISIS Facility, Rutherford Appleton Laboratory, Chilton, Didcot, OX11 0QX, U.K.}

\date{\today}

\begin{abstract}
The complex structure and magnetism of Pr$_{2-x}$Bi$_x$Ru$_2$O$_7$  was investigated by neutron scattering and EXAFS. Pr has an approximate doublet ground-state and the first excited state is a singlet. This overall crystal field level scheme is similar to metallic Pr$_2$Ir$_2$O$_7$, which is also reported here. While the B-site (Ru)  is well ordered throughout, this is not the case for the A-site (Pr/Bi). A distribution of the Pr-O2 bond length indicates the Pr environment  is not uniform even for $x=0$. The Bi environment is highly disordered ostensibly due to the 6s lone pairs on Bi$^{3+}$. Correspondingly we find the non-Kramers doublet ground state degeneracy otherwise anticipated for Pr in the pyrochlore structure is lifted so as to produce a quadrupolar singlet ground state with a spatially varying energy gap. For $x=0$, below T$_N$, the Ru sublattice orders antiferromagnetically, with propagation vector \textbf{k}= (0,0,0), as for Y$_2$Ru$_2$O$_7$. No ordering associated with the Pr sublattice is observed down to 100 mK. The low energy magnetic response of Pr$_{2-x}$Bi$_x$Ru$_2$O$_7$  features a broad spectrum of magnetic excitations associated with  inhomogeneous splitting of the Pr quasi-doublet ground state. For $x=0$ ($x=0.97$) the spectrum is temperature dependent (independent). It appears disorder associated with Bi alloying enhances the inhomogeneous  Pr crystal field level splitting so that inter-site interactions become irrelevant for $x=0.97$. The structural complexity for the A-site may be reflected in the hysteretic uniform magnetization of B-site ruthenium in the N\'{e}el phase.  
\end{abstract}


\maketitle

\section{Introduction}
In pyrochlore materials, with the general formula A$_2$B$_2$O$_7$, the A and B site ions form an interpenetrating network of corner-sharing tetrahedra.~\cite{pyrochlore_greedan} When populated by magnetic ions with nearest neighbor antiferromagnetic (AFM) interactions these materials display anomalous frustrated magnetism.~\cite{pyrochlore_anderson} For classical spins and nearest neighbor interactions the ground-state is a manifold characterized by zero magnetization on every tetrahedron.~\cite{pyrochlore_moessner}  Correspondingly pyrochlore magnets remain paramagnetic to much lower temperatures than their Curie-Weiss temperature ($|\Theta_{CW}|$) where the collective properties are determined by longer range or anisotropic interactions and thermal and/or quantum fluctuations. The resulting low temperature phases include spin-glasses, spin-liquids, and magneto-elastically induced N\'{e}el order.~\cite{pyrochlore_gardner1, pyrochlore_gardner2, pyrochlore_machida, pyrochlore_Tchernyshyov} 

While the majority of pyrochlore magnets are insulators, an interplay between magnetism and strong electron correlations can occur when the B-site is a $4d$- or $5d$-ion  resulting in a metal-to-insulator transitions (MIT), heavy fermion behavior and even superconductivity.~\cite{pyrochlore_sleight, pyrochlore_takeda, pyrochlore_tachibana,  pyrochlore_sakai} The ruthenium pyrochlores for example display a variety of ground states near a correlation induced MIT.~\cite{R2Ru2O7_1, R2Ru2O7_2, Y2Ru2O7, Dy2Ru2O7,Er2Ru2O7, Ho2Ru2O7, Bi2Ru2O7, Tl2Ru2O7} Their electronic bandwidths are strongly influenced by the Ru-O-Ru bond angle, which in turn is controlled by the ionic radius of the A$^{3+}$ ion.~\cite{A2Ru2O7} Thus R$_2$Ru$_2$O$_7$ (R = Y, rare earths) are insulating and show long-range magnetic order while Bi$_2$Ru$_2$O$_7$ is a Pauli paramagnet. Tl$_2$Ru$_2$O$_7$ is a metal at room temperature and a spin-singlet insulator below 120 K. Bulk measurements show antiferromagnetic ordering of the Ru sublattice for the insulating compounds with a critical temperature, T$_N$, that decreases monotonically from 160 K for Pr to 81 K for Yb; consistent with the lanthanide contraction.~\cite{Y2Ru2O7} The temperature and energy scale of the magnetic interactions on the rare earth sublattice is an order of magnitude lower than for the transition metal B-site.~\cite{R2Ru2O7_1, R2Ru2O7_2,Er2Ru2O7, Ho2Ru2O7,Dy2Ru2O7} 

In this paper we examine the structure and magnetism of Pr$_{2-x}$Bi$_x$Ru$_2$O$_7$ solid solutions.\cite{prbiru2o7_PRL} Bulk measurements show substituting Bi$^{3+}$ for Pr$^{3+}$  drives the system from  an antiferromagnetic insulator ($x= 0$) to a Pauli paramagnetic metal ($x=2$). While this transition has been observed in other Ru pyrochlores, in Pr$_{2-x}$Bi$_x$Ru$_2$O$_7$ the low-$T$ specific heat is greatly enhanced, reminiscent of what is observed in non-Fermi-liquid and heavy fermion systems. 

In both Pr$_{2-x}$Bi$_x$Ru$_2$O$_7$ and Pr$_2$Ir$_2$O$_7$, we find a quasi-doublet ground-state for Pr and a singlet excited state with analogous wave functions. In previous inelastic neutron scattering experiments we showed the enhanced low-$T$ specific heat and heavy fermion-like properties are actually a consequence of a static inhomogeneous splitting of the non-Kramers Pr$^{3+}$ ground-state doublet. Here we show that even without Bi substitution in $\rm Pr_2Ru_2O_7$, the ground state degeneracy anticipated for non-Kramers praseodymium in the pyrochlore lattice A-site is lifted. This is evidence of a local structural distortion that breaks the three fold rotation axis. Through EXAFS, we then provide direct structural evidence for a distribution of coordinating environments for praseodymium in  $\rm Pr_2Ru_2O_7$. While the B-site (Ru) environment remains well ordered throughout the series, the A-site becomes progressively disordered with increasing $x$, primarily near bismuth. 

For $x= 0$, below T$_N$, the Ru sublattice orders in a similar arrangement as for Y$_2$Ru$_2$O$_7$  so this order does not appear to be influenced by the Pr rare earth anisotropy.~\cite{Y2Ru2O7} For the Pr sublattice however, no order is detected by diffraction down to 1.5 K for any $x$. The specific heat has a Schottky-like anomaly centered at 3 K but no further anomalies associated with magnetic ordering at least down to 0.1 K. Probed by inelastic neutron scattering, the low energy magnetic excitation spectrum of Pr$_{2-x}$Bi$_x$Ru$_2$O$_7$ shows the corresponding mode of excitation. In the temperature dependence of the excitation spectrum we provide evidence for collective effects from Pr-Pr interactions for $\rm Pr_2Ru_2O_7$, these are shown to vanish for  $x= 0.97$ where the temperature dependence of the inelastic scattering can be described by inhomogenous single ion physics. 

\section{Experimental Techniques}
Powdered samples of Pr$_{2-x}$Bi$_x$O$_7$ ($x=0$, $x=0.97$ and $x=2$) and Pr$_2$Ir$_2$O$_7$ were synthesized using the solid state reaction method. For the Ru containing samples, mixtures of Pr$_2$O$_3$, Bi$_2$O$_3$ and RuO$_2$ in proper molar ratios were pre-reacted at 850 $^\circ$C for 15 h in air and then ground and pressed into pellets. In the case of Pr$_2$Ir$_2$O$_7$ mixtures of Pr$_6$O$_{11}$ and IrO$_3$ in proper molar ratios were pressed into pellets. The pellets were subsequently sintered at 1000-1200 $^\circ$C in air with intermediate grindings. All samples were characterized by powder X-ray diffraction. These measurements showed the samples all adopt the cubic pyrochlore structure and are single phase, except for the $x=0$ sample which contained 3.38(5) wt \% of unreacted RuO$_2$. Detailed bulk measurements on these samples have been reported elsewhere.~\cite{prbiru2o7_PRL,pr2ir2o7_JPCS} 

For the heat capacity measurement $\rm Pr_2Ru_2O_7$ powder was thoroughly mixed with silver powder 50\% by weight and cold pressed into a solid pellet to achieve adequate thermal conductivity for thermal equilibration. Data were collected down to 90 mK with the adiabatic relaxation method using a commercial Physical Property Measurement System (PPMS) Dilution Refrigerator. The specific heat capacity of $\rm Pr_2Ru_2O_7$ was obtained by subtracting the measured  specific heat capacity of silver  from the measured total heat capacity.~\cite{Ag_Cp}

EXAFS studies were carried out at the Stanford Synchrotron Radiation Lightsource (SSRL) for all the metal edges in $x= 0.97$ and the pure end compounds $x=0$ and $x=2$.  Transmission mode EXAFS data were collected for the Pr L$_{\rm III}$-edge (5964 eV), Bi L$_{\rm III}$-edge (13419 eV), and Ru K-edge (22,117 eV). We used a Si (220) double monochromator for the Ru edge and Si (111) crystals for the Bi and Pr L$_{\rm III}$ edges. The slit height was 0.5 mm, giving energy resolutions of 1 eV for the Pr L$_{\rm III}$-edge, and $\sim$2.7 eV for the Bi L$_{\rm III}$- and the Ru K-edge. The monochromator was detuned 50\% for the Pr and Bi L$_{\rm III}$ edges and 30\% for the Ru K-edge to minimize harmonics. EXAFS samples were prepared by first brushing fine powder ($\leq$ 5 $\mu$m) onto scotch tape; two pieces of tape were then pressed together (double layer) to encapsulate the powder.  For the Ru edge we used 7, 9, and 15 double layers for $x=2$, $x=0.97$ and $x=0$ respectively. 3 double layers were used for the Bi L$_{\rm III}$ edges, and 2 double layers for the Pr L$_{\rm III}$-edges.

Standard procedures were used to reduce the EXAFS data.~\cite{RSXAP} First a pre-edge subtraction was done to remove absorption from other atoms; this yields the absorption edge of interest, $\mu_{edge}$. Then a spline was fit through the data above the edge to obtain an estimate of the absorption, $\mu_0$, with no photoelectron backscattering. Next the EXAFS oscillations, $\chi(E)$, were obtained from $\mu_{edge}$ = $\mu_0$(1+$\chi(E)$), and $\chi(E)$ converted to $\chi(k)$ using $\hbar^2$ $k^2$/2m = $E-E_0$, where $E_0$ is the absorption edge energy. Finally $k^n\chi$(k) (usually n = 1-3) was Fourier transformed (FT) into r-space, where peaks in the FT correspond to various neighboring shells about the absorbing atom.

Powder neutron diffraction data were collected on the $x=0$ sample at the National Institute of Standards and Technology in Gaithersburg, Maryland (BT1) and the ISIS Facility, Rutherford Appleton Laboratory, UK (HRPD). For the BT1 experiment a 10 g  sample was sealed in a vanadium container with length 50 mm and diameter 10.8 mm and  temperature was controlled in a He cryostat. A Ge (311) monochromator with a 90$^\circ$ take-off angle ($\lambda$= 2.079 \AA) and 15 minutes of arc in-pile collimation were used. Data sets were collected for temperatures between 1.5~K and 180~K and scattering angle $2\theta$ from  3-168$^\circ$  with a step size of 0.05$^\circ$. For the HRPD experiment 10 g of sample was placed in a vanadium container 15 mm$\times$20 mm$\times$10 mm (h$\times$w$\times$d) within a He cryostat. Data sets were collected for temperatures between 2~K and 300~K using 10 ms  to 110 ms chopper settings. Rietveld analysis of the neutron powder diffraction patterns was performed using the Fullprof software package.~\cite{fullprof}

Inelastic neutron scattering experiments were carried out at the ISIS Facility, Rutherford Appleton Laboratory, UK. High energy data were collected on all samples using the time-of-flight (TOF) spectrometer HET. Additional low energy data were collected on the $x=0$ sample using the IRIS spectrometer.~\cite{IRIS_1, IRIS_2} For the HET experiment the samples were loaded in an Al sachet and the total mass of sample in the beam was 19.2 g for $x=0$, 24.9 g for $x=0.97$, 22.55 g for $x=2$ and 14.086 g for Pr$_2$Ir$_2$O$_7$. The samples were loaded into a top loading closed cycle He refrigerator. Incident energies of $E_i$= 35 and 160 meV were used for this experiment, the full width at half maximum (FWHM) energy resolution at the elastic line was 1.4 meV and 7 meV respectively. Data were collected at 5~K and 200~K. More details on normalization and the correction for the phonon contribution to the scattering data will be given below. Crystal Field (CF) analysis of the data was performed using the FOCUS program.~\cite{focus} 

The IRIS experiment was carried out on 15 g of $\rm Pr_2Ru_2O_7$. The sample was held in a sealed 2 mm double walled Al can with diameter 23 mm and height 53 mm and loaded into a He cryostat. Bandwidth disk choppers selected an incident spectrum from 1.35 meV to 4.6 meV above $E_f = 1.847$~meV pulsed at 25 Hz and a backscattering pyrolytic graphite analyzer bank with a 25 K Be filter selected the final energy, $E_f$. The FWHM elastic energy resolution was 17.5 $\mu$eV. Data were collected over a temperature range from 1.5~K to 200~K.

The magnetic neutron scattering cross-section of a powder sample at wave vector transfer $Q$ and energy transfer $\hbar\omega$ can be written as~\cite{neutron-cross}
\begin{eqnarray}
\frac{d^2\sigma}{d\Omega dE^{\prime}}&=&
\frac{k_f}{k_i}(\gamma r_0)^2 |\frac{g}{2} F({\mathbf Q})|^2 
\int \frac{d\Omega_{\hat{Q}}}{4\pi} \nonumber\\
 & &\times \sum_{\alpha,\beta}(\delta_{\alpha\beta}-\hat{Q}_{\alpha}\hat{Q}_{\beta}){\cal S}^{\alpha\beta}({\mathbf Q}, \omega)
\label{eq:n-cross}
\end{eqnarray}
where $\gamma$= -1.913 and $g$ are the spectroscopic g-factors of the neutron and the magnetic ion respectively, $r_0=e^2/m_ec^2= 2.82$~fm is the classical electron radius, and $F({\bf Q})$ is the magnetic form factor.~\cite{form_fact} The dynamic spin correlation function ${\cal S}^{\alpha\beta}({\mathbf Q}, \omega)$ is given by
\begin{eqnarray}
{\cal S}^{\alpha\beta}({\mathbf Q}, \omega)&=&\frac{1}{2\pi\hbar} \int dt e^{i\omega t} \nonumber\\
& &\times\frac{1}{N} \sum_{ij}\langle S_{i}^{\alpha} (t)S_{j}^{\beta} (0)\rangle e^{-i{\bf Q}\cdot ({\bf R}_i-{\bf R}_j)}.
\label{eq:dyn_spin_cor}
\end{eqnarray}
Here $N$ is the number of formula units encompassed in each of the double summations. 
${\cal S}^{\alpha\beta}({\mathbf Q}, \omega)$ can be related to the generalized susceptibility through the fluctuation-dissipation theorem~\cite{neutron-cross}
\begin{equation}
{\cal S}({\bf Q},\omega )=
\frac{1}{1-e^{-\beta\hbar\omega }}
\frac{\chi^{\prime\prime}({\bf Q},\omega )}{\pi (g\mu_B)^2}
\end{equation}
where $\beta=1/k_BT$ and $\chi^{\prime\prime}$ denotes the imaginary part of the generalized susceptibility. In a system that contains magnetic rare earth ions where the free ion ground-state $J$ multiplet is split due to the effect of a CF, the generalized susceptibility can be calculated from the eigenfunctions and energies of the CF Hamiltonian. For a cubic material where inter-site interactions are treated in the Random Phase Approximation, the generalized susceptibility per formula unit can be expressed as 
\begin{eqnarray}
\chi({\bf Q}, \omega)= \frac{n}{3} \sum_{\alpha} \frac{\chi_0^{\alpha\alpha}(\omega)}{1-\lambda({\bf Q})\chi_0^{\alpha\alpha}(\omega)}
\label{eq:gen_suscept}
\end{eqnarray}
where $n$ is the number of magnetic ions per formula unit, $\lambda({\bf Q})$ is the exchange interaction and $\chi_0^{\alpha\beta}(\omega)$ is the single site susceptibility, which can be expressed as follows~\cite{suscept_cal}
\begin{eqnarray}
\lefteqn{\chi_0^{\alpha\beta}(\omega)=} \nonumber\\
 & &(g\mu_B)^2\lim_{\epsilon \rightarrow 0^+} [ 
\sum_{\stackrel{p,q}{\scriptstyle E_p\neq E_q}} 
\frac{<p|J_{\alpha}|q><q|J_{\beta}|p>}{E_p-E_q-\omega-\mathrm{i}\epsilon} (n_q-n_p) \nonumber \\
 & & +\frac{1}{k_BT}\frac{\epsilon}{\epsilon- \mathrm{i}\omega}(
\sum_{\stackrel{p,q}{\scriptstyle E_p= E_q}} <p|J_{\alpha}|q><q|J_{\beta}|p> n_p \nonumber \\
& & -<J_{\alpha}><J_{\beta}>)].
\end{eqnarray}
Here $J_{\alpha}$ indicates a carterian component ($\alpha= x, y, z$) of the angular momentum operator, $|p>$ and $E_p$ are the eigenfunctions and energies of the crystal field Hamiltonian ${\cal H}_{CF}$, and $n_p$ is the thermal population factor. For Pr$^{3+}$ ions in the pyrochlore structure we expect the 9-fold degenerate free ion ground-state $J$ multiplet $^4H_3$ to split into 3 doublets and 3 singlets under the effect of the $D_{3d}$ symmetric ${\cal H}_{CF}$. Choosing [111] as the quantization axis, the single-ion crystal field Hamiltonian takes the form:
\begin{equation}
{\cal H}_{CF}= B^0_2O^0_4+B^0_4O^0_4+B^3_4O^3_4+B^0_6O^0_6+B^3_6O^3_6+B^6_6O^6_6 \label{eq:CF_Ham}
\end{equation}
where $B^m_n$ are the CF parameters and $O^m_n$ are Stevens operator equivalents of the CF tensor operators as discussed by Hutchings.~\cite{hutchings} Here the CF interaction in the LS coupling scheme is treated as a perturbation within the ground-state $J$ multiplet only. 

The following corrections were applied to the neutron counts in the time histograms collected on both HET and IRIS. First a time-independent background measured for $\hbar\omega\approx$ -$E_f$ was subtracted. Then the data were scaled to the relevant count rate in a pre-sample monitor and finally converted into $\hbar\omega$ histograms. This procedure gives $\hbar\omega$-dependent data $I(Q,\hbar\omega)$, which are related to the scattering cross-section through convolution with a resolution function, as follows:
\begin{eqnarray}
I(Q,\hbar\omega)&=&{\cal C}N \int dQ^{\prime} \hbar d\omega^{\prime} R_{Q\omega}(Q-Q^{\prime}, \omega-\omega^{\prime}) \nonumber\\
 & &\times \frac{k_i}{k_f} \frac{d^2\sigma}{d\Omega dE^{\prime}}(Q^{\prime}, \omega^{\prime}),
\end{eqnarray}
where $N$ is the number of formula units in the sample, and ${\cal C}$ is the spectrometer constant. The instrumental resolution function $R_{Q\omega}$ is assumed to be unity normalised:
\begin{equation}
1\equiv\int R_{Q\omega}(Q-Q^{\prime}, \omega-\omega^{\prime}) dQ^{\prime} d\omega^{\prime}
\end{equation}

The normalized intensity $\tilde{I}(Q,\hbar\omega)$ is related to the measured intensity as follows
\begin{equation}
\tilde{I}(Q,\hbar\omega)= \frac{I(Q,\hbar\omega)}{{\cal C}N}.
\label{eq:norm_int}
\end{equation}
Thus $\tilde{I}(Q,\hbar\omega)$ is the resolution smeared partial differential scattering cross section per formula unit which we express 
in absolute units of mbarn sr$^{-1}$ meV$^{-1}$ f.u.$^{-1}$. 

For the HET experiment,  ${\cal C}N$ was determine by measuring the incoherent scattering from a standard flat vanadium slab sample for each of the chosen incident energies. For the IRIS experiment, ${\cal C}N$ was determined from Bragg scattering through a method that has been described elsewhere.~\cite{bragg_norm} These procedures yield absolute measurements of $\tilde{I}(Q,\hbar\omega)$ to an overall scale accuracy of 20 $\%$.

\section{Results}

\subsection{EXAFS measurements}
To investigate possible distortions in the local structure  we  carried out EXAFS measurements. Here we present the main results of the EXAFS analysis. 

\begin{figure}[t]
\includegraphics[width= 8.5cm,clip]{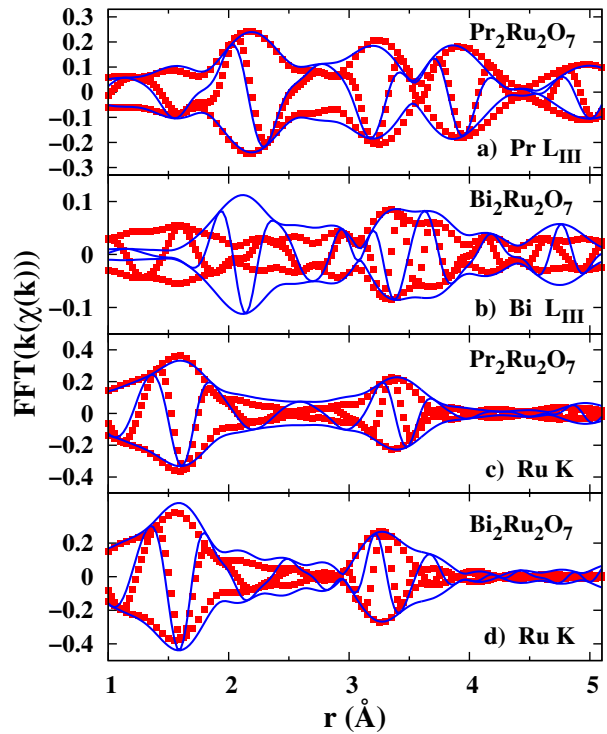}
\caption{EXAFS $r$-space data at 4K for a) the Pr L$_{\rm III}\rm{-}$, b) the Bi L$_{\rm III}$-edge, and c), d) the Ru K-edges for $x=0$ and $x=2$ respectively; the data are shown as solid squares. The solid line in each panel is a simulation (not a fit) using the program FEFF8.2, the ordered pyrochlore structure, and a global broadening parameter (0.07 \AA{} for Pr and 0.08 \AA{} for Bi). For the Ru edge data, we used 0.0725 \AA{} for $x=0$ and 0.05 \AA{} for $x=2$. The largest deviations are for the Bi L$_{\rm III}$ edge at the first neighbor O peak near 2 \AA{}  (a sum of Bi-O1 and Bi-O2 contributions). The FT ranges are Pr L$_{\rm III}$, 3.5-10 \AA$^{-1}$; Ru K, 4.5-14 \AA$^{-1}$; and Bi L$_{\rm III}$, 4-14 \AA$^{-1}$; with a Gaussian rounding of the transform window by 0.3 \AA$^{-1}$.  In this and subsequent $r$-space plots, the fast oscillation is the real part R of the FT while the envelop function is $\pm$ $\sqrt{R^2 + I^2}$ where I is the imaginary part of the FT.}\label{r-space-pure}
\end{figure}

In Figure~\ref{r-space-pure}a,b we first show the experimental $r$-space plots
(FT$ k\chi$(k)) at T=4 K (solid squares) for the A-site atoms (Pr and Bi) in
the end compounds $x=0$ and $x=2$. The first peak in each scan is the metal-O
peak, near 2.0-2.5 \AA; a sum of Pr-O2 (2.243 \AA) and Pr-O1 (2.546 \AA) for the Pr
L$_{\rm III}$ edge, and a sum of Bi-O2 (2.228 \AA) and Bi-O1 (2.538 \AA) peaks
for the Bi L$_{\rm III}$ edge. The next peak (near 3.3 \AA) is a combination of metal-metal
peaks, i.e. for $x=0$ at the Pr edge it would be a sum of Pr-Ru and Pr-Pr
peaks. 

Note the peaks in the EXAFS spectra are shifted to lower $r$ compared to
the actual distances, by a well known phase factor. For example the two Pr-O
peaks in the EXAFS spectra of Fig.~\ref{r-space-pure}(a) are located at $\sim$ 1.8 and 2.1 \AA{} if plotted
separately - a shift of roughly -0.45 \AA. For these peaks, the $r$-space phase
(real part of the transform) of the peak for the shortest Pr-O2 distance (2
neighbours) is nearly out of phase with that for the longer distance Pr-O1 peak
(6 neighbours), leading to a dip in the spectra at 1.8 \AA. 

Figure~\ref{r-space-pure}c, d shows the corresponding Ru K-edge data for the two pure samples (solid squares); c) $x=0$, d) $x=2$. The two data sets are very similar. For Ru, there is only one nearby O neighbor (O1), while the second peak is a sum of Ru-Ru and either Ru-Pr (Fig.~\ref{r-space-pure}(c)) or Ru-Bi (Fig.~\ref{r-space-pure}(d)). 

A quick evaluation of the data is obtained by simulating the EXAFS $r$-space data using the program FEFF8.2, the known ordered pyrochlore structure, and a global broadening (pair distribution width, $\sigma$) for all peaks.~\cite{FEFF8, Avdeev02} These simulations are shown as solid lines in Figure \ref{r-space-pure}, with $\sigma$ = 0.07 and 0.08 \AA{} for the Pr and Bi edges respectively, and 0.0725 and 0.05 \AA{} for the Ru K-edge data for $x=0$ and $x=2$. Note that the only parameter adjusted here was the global broadening.

This modeling shows the environment about Pr is relatively well ordered (some disorder for Pr-O2
and Pr-Pr is discussed later) whereas that about Bi is highly disordered -- the Bi-O1 peak which should occur near 2.1 \AA{} in Figure~\ref{r-space-pure}(b), is strongly suppressed. The weak peak near 1.7-1.8 \AA {} is consistent with the small Bi-O2 peak (two O neighbors located inside the Bi/Pr tetrahedra). For the Ru edge data, the Ru-O1 peak near 1.6 \AA{} is also well ordered, surprisingly, even for the Bi sample which has significant disorder of the Bi-O peak. Consequently since the Ru-O1 peak has little disorder while the Bi-O1 peak is disordered, Bi must be displaced from the usual A-site position, in a direction perpendicular to the Bi-O2 axis. This leads to a small distortion of the Bi-O2 peak.

\begin{figure}
\includegraphics[width= 8.5cm,clip]{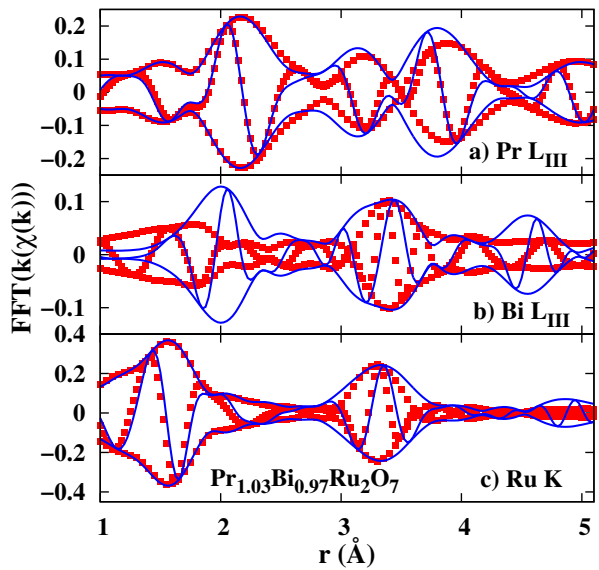}
\caption{EXAFS $r$-space data at 4 K for $x=0.97$: a) the Pr L$_{\rm III}\rm{-}$, b) the Bi L$_{\rm III}\rm{-}$ and c) the Ru K-edge. The solid squares are the experimental data while the solid lines are simulations, again using the
program FEFF8.2, and global broadening parameters (0.06 \AA{} for Pr, 0.065 \AA{} for Bi, and 0.05 \AA{} for Ru). The largest deviation again occurs for the Bi L$_{\rm III}$ data (panel b) at the O-peak near 2 \AA. Same FT ranges as in Figure~\ref{r-space-pure}.} \label{r-space-mixed} 
\end{figure}

Next we consider disorder in the mixed sample,
$x=0.97$. In Figure~\ref{r-space-mixed} we show the
$r$-space plots for the three edges (Pr and Bi L$_{\rm III}$-, and Ru K-edge) at 4K. For each plot the first peak corresponds to the nearest neighbor O shell; both the combined Pr-O1/Pr-O2 peak in Fig.~\ref{r-space-mixed}a and the Ru-O1 peak in Fig.~\ref{r-space-mixed}c have a large amplitude
indicating relatively little disorder. In contrast, the Bi-O peak (Bi-O1 and Bi-O2) in panel b) is suppressed, mostly at the position for Bi-O1, similar to the results for the pure end compound $x=2$.  This indicates that Bi is also displaced perpendicular to the Bi-O2 axis in the mixed compound and by a comparable amount.

We find similar results at higher temperatures, the Ru-O and most Pr-O peaks are generally well ordered while the Bi-O peak is strongly suppressed, indicating significant disorder. We also find that the Pr-Ru, Ru-Pr and Ru-Ru second neighbor peaks are reasonably ordered, but when Bi is present, peaks that include Bi second neighbors (e.g. Pr-Bi or Ru-Bi) also have disorder.

\subsection{Crystal field measurements}
To determine the  crystal field level scheme and the relevant low energy spin degrees of freedom we carried out high energy inelastic neutron scattering measurements on HET. Three methods were used to determine and then subtract the nonmagnetic phonon contribution to the data. For Pr$_{2-x}$Bi$_x$O$_7$ the scaling and direct subtraction methods were used, while for Pr$_2$Ir$_2$O$_7$ the DISCUS package was used.~\cite{discus} In all three methods the nonmagnetic contribution to the neutron scattering at low angles was subtracted by scaling the spectrum measured at high wave vectors where the magnetic response is negligible. 

In the so-called scaling method, the nonmagnetic contribution at low angles ($2\theta\approx$ 19$^\circ$) is estimated from the measured scattering at high scattering angles ($2\theta\approx$ 135$^\circ$) by using an energy dependent scaling factor determined from direct measurements on the Pauli paramagnetic compound $x=2$. This procedure is based on two assumptions. First, the magnetic scattering intensity, which is proportional to the square of the Pr$^{3+}$ magnetic form factor, is negligible in the high angle scattering data. Second, the energy dependent ratio between phonon scattering at low and high scattering angles is the same for all four compounds. 

The second method employs a direct subtraction method to estimate phonon contributions using the $x= 2$ compound as the phonon blank material, after accounting for the difference in the total scattering cross section $\sigma$($x= 0.97$)$\sim$ 0.9$\sigma$($x= 2$) and $\sigma$($x= 0$)$\sim$ 0.8$\sigma$($x= 2$). Both methods produced a very similar magnetic response though with less statistical error for the first scaling method. As a result all the data collected on Pr$_{2-x}$Bi$_x$O$_7$ were analyzed using the first method. In the case of Pr$_2$Ir$_2$O$_7$ the DISCUS package uses a Monte Carlo method to calculate a wave-vector-dependent scaling factor, which showed no significant energy dependence.

\begin{figure}
\includegraphics[width= 8.5cm]{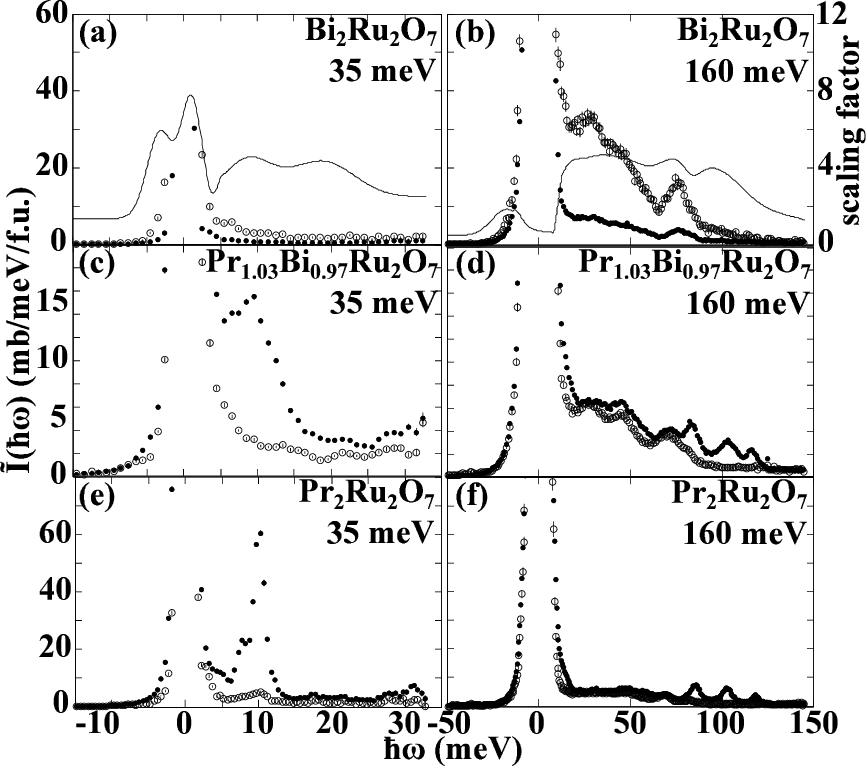}
\caption{Spectra of Pr$_{2-x}$Bi$_{x}$Ru$_2$O$_7$ for $x= 2$ (a, b), $x= 0.97$ (c, d) and $x= 0$ (e, f) taken at 5 K and $E_i$= 35 meV and 160 meV. The spectra taken at low scattering angles ($2\theta$= 19$^\circ$, Q$_{el}$= 1.36 \AA$^{-1}$ and 2.91 \AA$^{-1}$ respectively) are shown as $\bullet$ whereas those taken at high scattering angles ($2\theta$= 135$^\circ$, Q$_{el}$= 7.62 \AA$^{-1}$ and 16.30 \AA$^{-1}$) are shown as $\circ$. The solid line in (a,b) shows the energy dependent scaling factor as determined from fits to the $x= 2$ data. For $x= 0.97$ and $x= 0$ the high angle data have been scaled using the energy dependent scaling factor to show the estimate of the nonmagnetic phonon contribution to the low angle angle data.\label{phsub}}
\end{figure}

\begin{figure}
\includegraphics[width= 8.5cm]{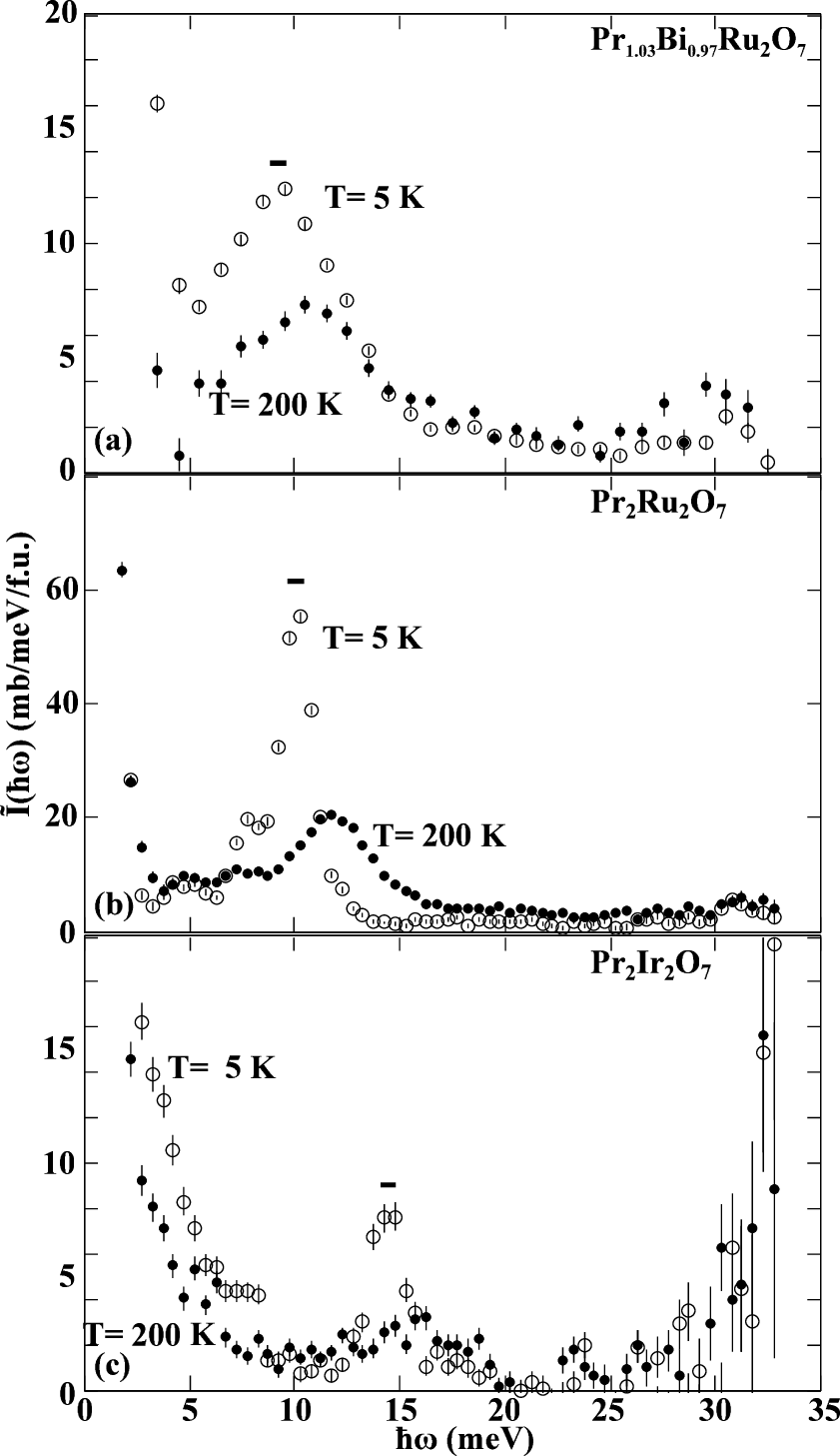}
\caption{The $E_i$= 35 meV magnetic response of Pr$_{2-x}$Bi$_{x}$Ru$_2$O$_7$ for $x= 0.97$ (a), $x= 0$ (b) and Pr$_2$Ir$_2$O$_7$ (c) at low scattering angles ($2\theta$= 19$^\circ$) at 5 K ($\circ$) and 200 K ($\bullet$) after subtracting off the nonmagnetic phonon background. The horizontal bar at 10 meV (15 meV for Pr$_2$Ir$_2$O$_7$) indicates the instrumental resolution at that energy transfer.\label{mr_35meV}}
\end{figure}

\begin{figure}
\includegraphics[width= 8.5cm]{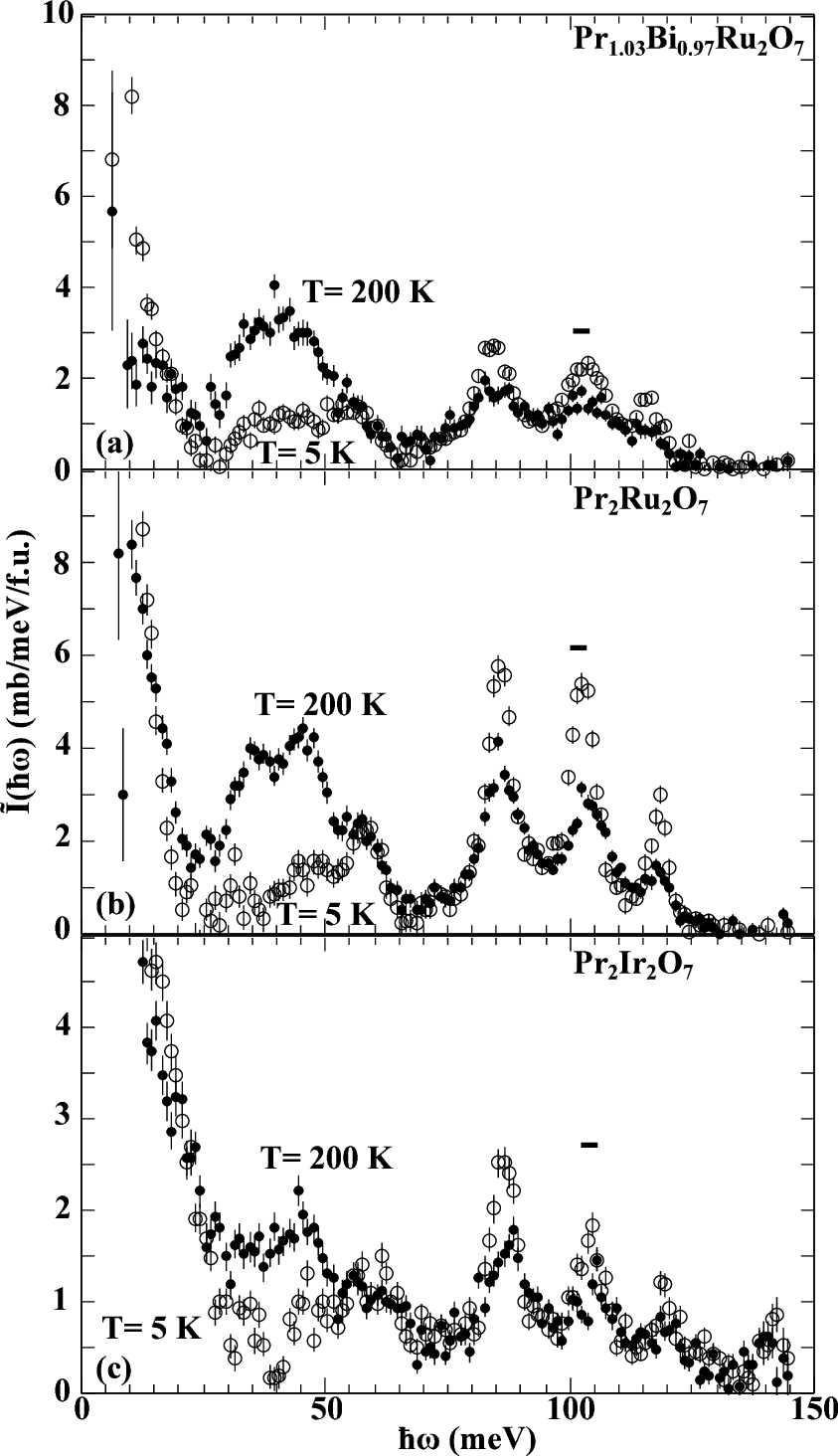}
\caption{The $E_i$= 160 meV magnetic response of Pr$_{2-x}$Bi$_{x}$Ru$_2$O$_7$ for $x= 0.97$ (a), $x= 0$ (b) and Pr$_2$Ir$_2$O$_7$ (c) at low scattering angles ($2\theta$= 19$^\circ$) at 5 K ($\circ$) and 200 K ($\bullet$) after subtracting off the nonmagnetic phonon background. The horizontal bar at 105 meV indicates the instrumental resolution at that energy transfer.\label{mr_160meV}}
\end{figure}

Figure~\ref{phsub} shows the total spectra for Pr$_{2-x}$Bi$_{x}$Ru$_2$O$_7$ with $x= 2$, $x= 0.97$ and $x= 0$ measured at 5 K with incident energies $E_i$= 35 meV and 160 meV for low ($2\theta$ $\approx$19$^\circ$) and high ($2\theta$ $\approx$135$^\circ$) scattering angles. For $x= 0.97$ and $x= 0$ the high angle spectra have been scaled down using the energy dependent scaling factor determined from the $x= 2$ data (solid lines in Figs.~\ref{phsub}(a) and~\ref{phsub}(b)). At low scattering angles, corresponding to Q$_{el}$= 1.36 \AA$^{-1}$ and 2.91 \AA$^{-1}$ respectively, the spectra from $= 0.97$ and $x= 0$ contain both magnetic and phonon  contributions. In the high scattering angles spectra, corresponding to Q$_{el}$= 7.62 \AA$^{-1}$ and 16.30 \AA$^{-1}$ respectively however, the magnetic contributions are small due to the very small form factor for Pr 4\textit{f} electrons at such large Q values. As shown in Figs.~\ref{phsub}(a) and~\ref{phsub}(b), the inelastic response of $x= 2$ shows three clear peaks due to one-phonon scattering at 30 meV, 45 meV and 75 meV. These features are reproduced in the scaled high angle scattering data of $x= 0.97$ and $x= 0$, indicating that the phonon scattering is indeed similar for all three compounds. This justifies use of the scaling method to estimate the phonon contribution to the low scattering angle spectra for the $x= 0$ and $x= 0.97$ samples. Figures~\ref{mr_35meV} and~\ref{mr_160meV} show  magnetic scattering from Pr$_{2-x}$Bi$_{x}$Ru$_2$O$_7$ and Pr$_2$Ir$_2$O$_7$, after subtracting the phonon and elastic scattering, at 5 K and 200 K.


The magnetic neutron scattering cross section for $\rm Pr_2Ru_2O_7$ at 5 K shows at least five magnetic excitations centered near  10 meV, 50 meV, 85 meV, 105 meV, and 116 meV energy transfer (Fig.~\ref{mr_35meV}(b) and~\ref{mr_160meV}(b)). Closer examination of the 5 K data shows the excitations near 10 meV and 50 meV are broadened or split. The remaining three high energy excitations take the form of isolated resolution limited peaks. 

Comparing the spectrum at 200 K with that at 5 K the following changes are observed upon warming; the strongest peak near 10 meV is broadened, decreases in intensity and shifts upward to 12 meV. A new broad double peak structure that resembles the broad 50 meV peak appears near 40 meV. The three peaks near 100 meV remain in place but loose intensity on warming.  

In the nominal $D_{3d}$ point group symmetry of the pyrochlore lattice, Pr$^{3+}$ has five CF  excitations. The broadening and splitting of the two lowest energy CF excitation may indicate an inhomogenous environment for praseodymium, something we also find evidence for in high resolution measurements that will be described subsequently. Thermal expansion as well as magneto-striction and dipole fields from Ru$^{4+}$ ordering at $T_N= 165$~K may be responsible for the modifications in the lowest energy CF excitations near upon heating to 200~K. Anomalous changes in crystal field excitations resulting from ruthenium spin ordering were for example previously document in Ho$_2$Ru$_2$O$_7$.\cite{Ho2Ru2O7} 

Thermal population of the 10 meV CF level for $T=200$~K enables excitations from that level to higher energy CF levels to which dipole transitions are allowed from the excited state. Thus heating can produce extra versions of higher excitations downshifted by $\approx 12$~meV, which is the energy of the first excited CF state at 200~K. We interpret the heating induced peak near 40~meV as resulting from this mechanism. This implies a finite dipole matrix element between the 12 meV and 50 meV CF levels. On the other hand the loss of intensity for the three upper CF transitions indicates the dipole matrix elements between the first excited state state and these three levels is small or even zero. 

The corresponding  5 K data for the  $x= 0.97$ sample also shows 5  excitations  (Fig.~\ref{mr_35meV}(a) and~\ref{mr_160meV}(a)).  We associate all of these magnetic peaks with Pr$^{3+}$ CF excitations. Much as for $\rm Pr_2Ru_2O_7$, there are four relatively sharp features centered at 9~meV, 83~meV, 103~meV and 116~meV  and a broad maximum near 50 meV. The FWHM of these excitations is however, a factor 3 larger than for $x= 0$, an effect  we may ascribe to alloying induced disorder in the electrostatic conditions for Pr$^{3+}$.~\cite{prbiru2o7_PRL} The effects of heating to 200 K are very similar to observations in $\rm Pr_2Ru_2O_7$. As the $x= 0.97$ sample has no magnetic phase transition down to 2 K, the similarity of the $x=0$ and $x=0.97$ data suggests ruthenium magnetic ordering does not have a significant effect on praseodymium here.  

The 5 K and 200 K CF spectra for Pr$_2$Ir$_2$O$_7$ are quite similar to the ruthenium based pyrochlores (Fig.~\ref{mr_35meV}(c) and~\ref{mr_160meV}(c)). There are again five energy levels; here lying at  14~meV, 58~meV, 86~meV, 104~meV, and 120~meV. The small maximum at the very top of the spectrum in Fig.~\ref{mr_160meV}(c) may be a result of incomplete background subtraction for this strongly absorbing sample, which also is seen to impact the top of spectrum for $E_i=35$~meV in Fig.~\ref{mr_35meV}(c).   In comparison to the other samples, the shifts are largest for the lower energy levels, which generally appear to respond more to the crystalline environment. The thermal effects in this sample, which as $\rm Pr_{1.03}Bi_{0.97}Ru_2O_7$ has no magnetic order on the transition metal site, are qualitatively similar  to both $\rm Pr_2Ru_2O_7$ and $\rm Pr_{1.03}Bi_{0.97}Ru_2O_7$. 

\subsection{Neutron powder diffraction measurements}\label{sec_diffraction}
\begin{figure}
	\includegraphics[width= 8.5cm]{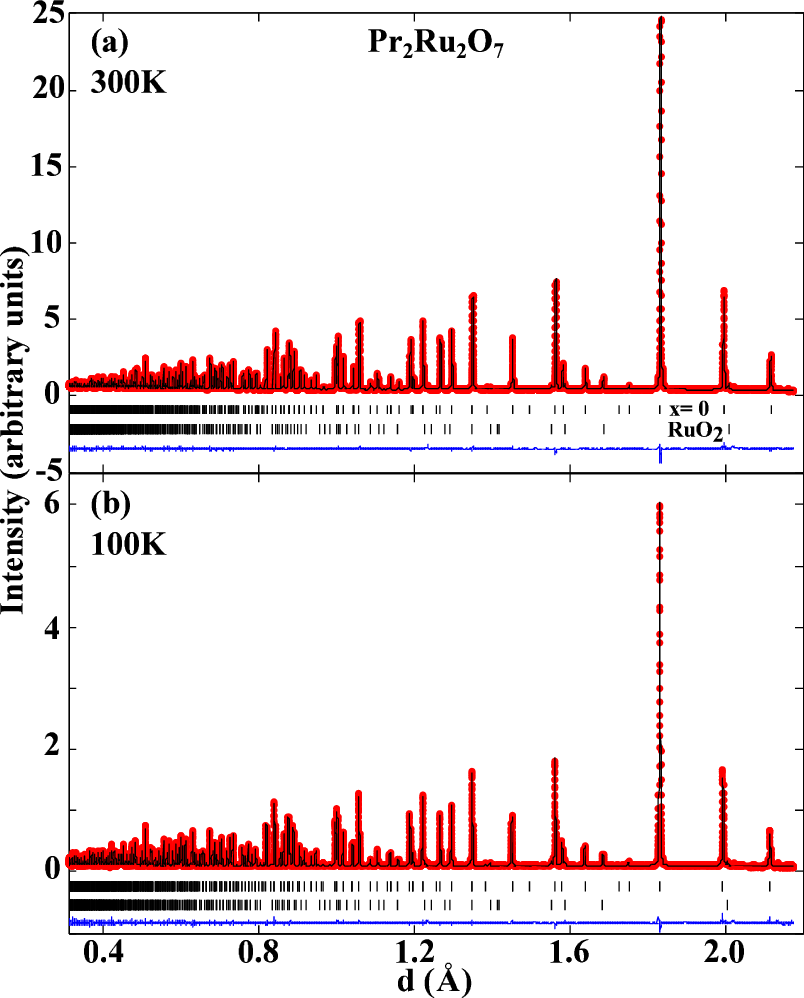}
	\caption{Neutron powder diffraction data of $x= 0$ at 300 K (a) and 100 K (b) collected on HRPD. The solid black line shows the Rietveld fit to the data, the residual of the fit (blue line) is shown at the bottom of the plot. The upper and lower tick marks indicate Bragg reflections coming from the crystal structure of the $x= 0$ and RuO$_2$ impurity phase respectively.\label{d_hrpd_fit}}
\end{figure}

\begin{figure}
\includegraphics[width= 8.5cm]{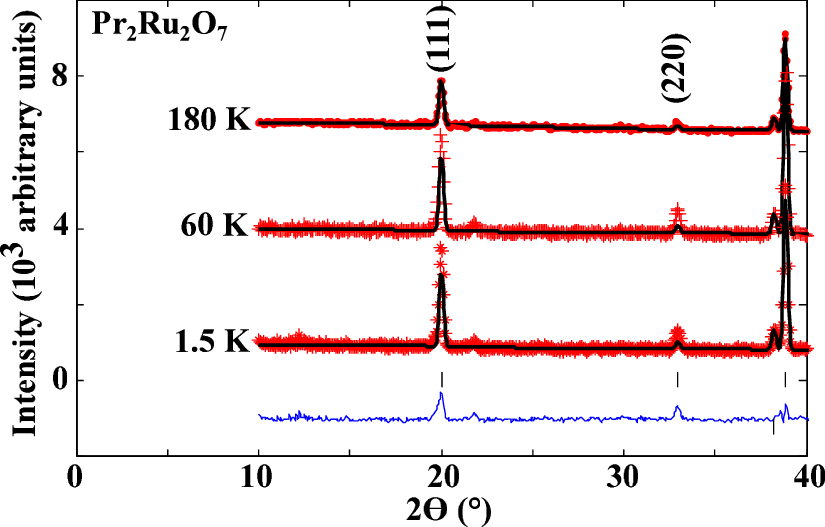}
\caption{Low angle part of the neutron powder diffraction profile of $x= 0$ as measured at 180 K ($\bullet$), 60 K (+) and 1.5 K (*) on BT1. The solid black lines show the Rietveld fit of the crystal structure (Table~\ref{tab:nucl_fit}) to the data, the residual of the 1.5 K fit (blue line) is shown at the bottom of the plot. The tick marks shown indicate Bragg reflections coming from the crystal structure of the $x= 0$ phase.\label{d_nucl_fit}}
\end{figure}

To determine  the magnetic ordering and potential structural distortions in $\rm Pr_2Ru_2O_7$, we carried out neutron diffraction studies on BT1 at NIST  and HRPD at ISIS. Figure~\ref{d_hrpd_fit} shows the Rietveld fits to the 300 K (T $>$ T$_N$) and 100 K (T $<$ T$_N$) data sets as collected on HRPD. These fits show $\rm Pr_2Ru_2O_7$ adopts the cubic pyrochlore structure and that the sample contained 3.46 wt \% of unreacted RuO$_2$. Fits of the crystal structure to the data collected below T$_N$ revealed no evidence, within the accuracy of the experiment, of a structural distortion associated with the magnetic phase transition. They do however show that below T$_N$, certain low angle reflections gain intensity that cannot be accounted for by the nuclear contributions alone (Figure~\ref{d_nucl_fit} and Table~\ref{tab:nucl_fit}).
\begin{table}
\caption{Refined structural parameters from fits to powder neutron diffraction profiles of the $x= 0$ sample collected at 180 K, 60 K and 1.5 K on BT1. The crystal structure is cubic with space group $Fd\bar{3}m$ with Pr located on $16(d)$ sites $(1/2,1/2,1/2)$, Ru located on $16(c)$ sites and O located on $48(f)$ (O1) and $8(b)$ (O2) sites $(x,1/8,1/8)$ and $(3/8,3/8,3/8)$ respectively.} \label{tab:nucl_fit}
	\centering
		\begin{tabular}{l|ccc}\hline \hline
			T (K) & 180 & 60 & 1.5 \\ \hline
			\textit{a} (\AA) & 10.36494(5) & 10.36048(4) & 10.36031(4) \\
			$x_{\mathrm{O1}}$ & 0.32919(8) & 0.32932(7) & 0.32929(6) \\
			$<u^2>$(Pr) (\AA$^2$)& 0.0094(5) & 0.0076(5) & 0.0075(5) \\
			$<u^2>$(Ru) (\AA$^2$)& 0.0032(4) & 0.0027(4) & 0.0027(3) \\
			$<u^2>$(O1) (\AA$^2$)& 0.0047(3) & 0.0046(3) & 0.0048(3) \\
			$<u^2>$(O2) (\AA$^2$)& 0.0047(6) & 0.0044(5) & 0.0041(5) \\
			R$_{wp}$ (\%) & 10.9 & 9.55 & 9.14 \\
			$\chi^2$ & 1.90 & 2.38 & 2.19 \\ \hline \hline
		\end{tabular} 
\end{table}

\begin{figure}
\includegraphics[width= 8.5cm]{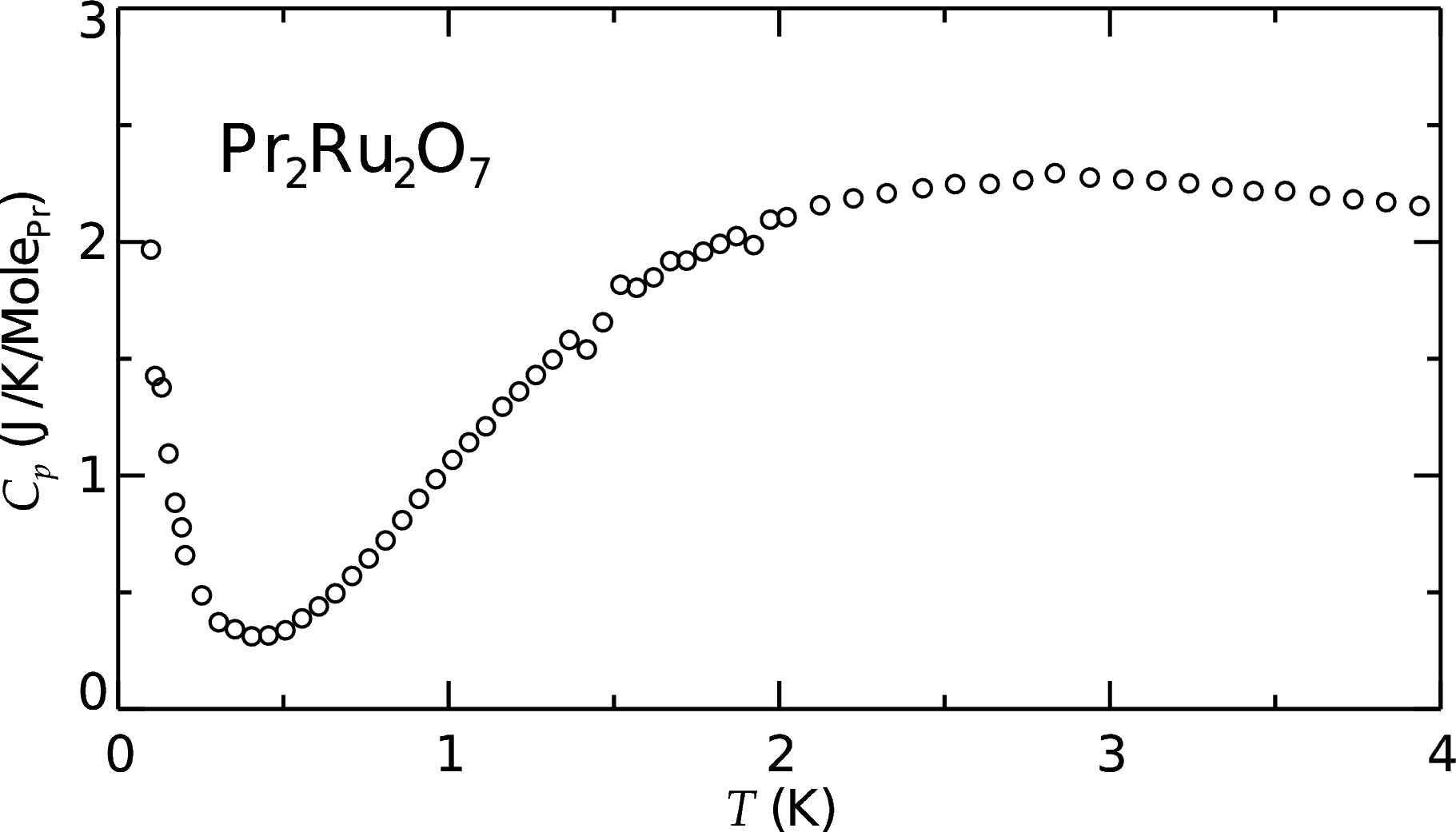}
\caption{Heat capacity of $x=0$ ($\circ$) normalized to the mole number of Pr.\label{Cp}}
\end{figure}

Figure~\ref{d_nucl_fit} shows fits (of the crystal structure) to the low angle part of the neutron powder diffraction profile of $x= 0$ measured above and below T$_N $. It can clearly be seen that for T $<$ T$_N$ there is additional intensity associated with the (111) and (220) reflections that can not be accounted for by nuclear contributions only. As the second phase RuO$_2$ is a Pauli paramagnet, this additional intensity must be due to  long range ordering of Ru dipole moments in $\rm Pr_2Ru_2O_7$. The enhanced (111) and (222) magnetic scattering resembles our results for Y$_2$Ru$_2$O$_7$, but differs from the structures observed in Ho$_2$Ru$_2$O$_7$ and Er$_2$Ru$_2$O$_7$.~\cite{Y2Ru2O7,Ho2Ru2O7,Er2Ru2O7} Down to 1.5 K we did not detect additional intensity that might be associated with ordering and/or freezing of the  Pr sublattice. This is consistent with our heat capacity measurements, which show there is no additional phase transition in the relevant low temperature range  (Figure~\ref{Cp}). 

\subsection{Low energy excitations}
\begin{figure}
	\includegraphics[width= 8.5cm]{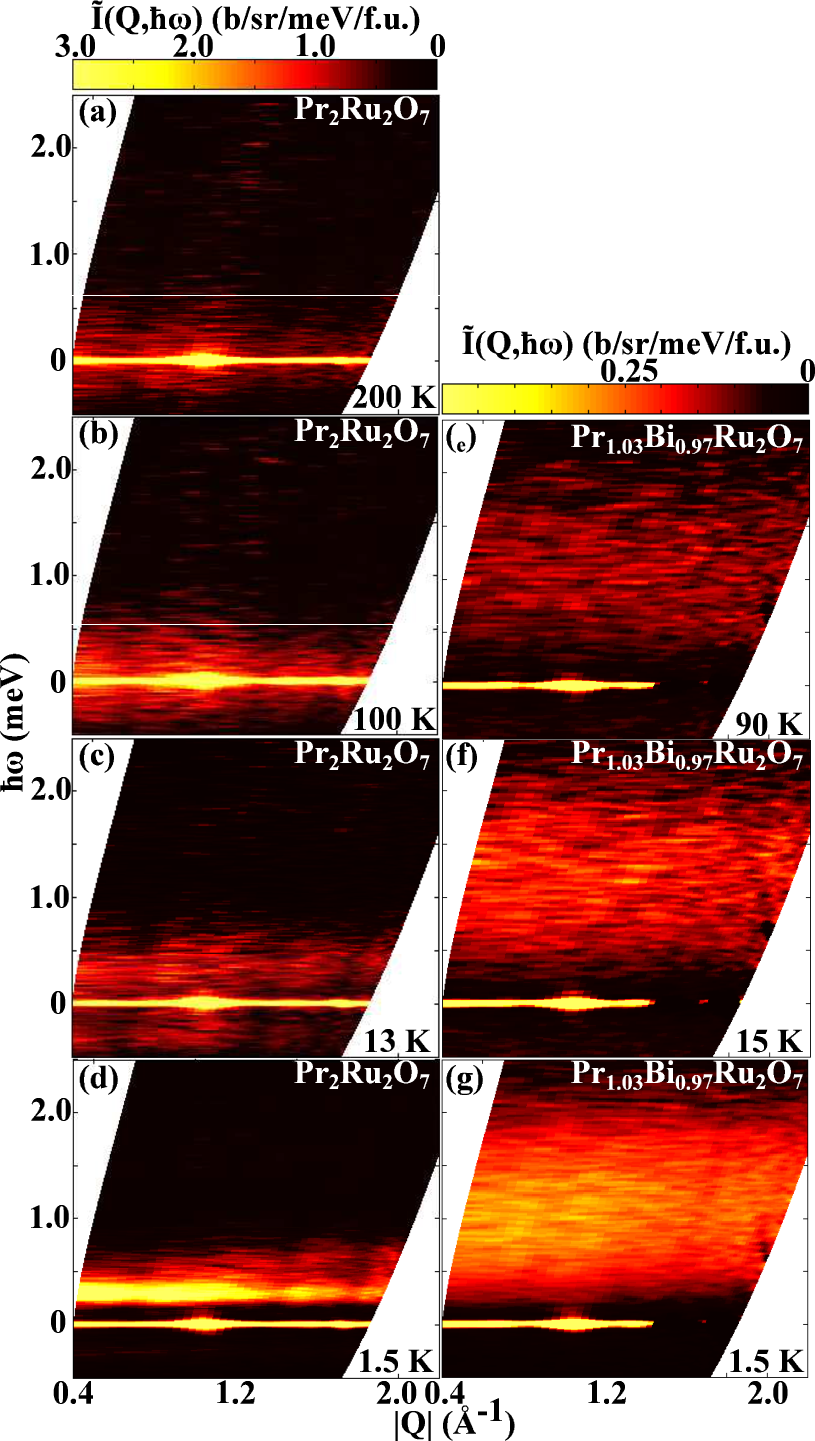}
	\caption{Low energy magnetic response, $\tilde{I}(Q,\hbar\omega)$, of $x= 0$ (at 200 K (a), 100 K (b), 13 K (c) and 1.5 K (d)) and $x= 0.97$ (at 90 K (a), 15 K (b) and 1.5 K (c)).~\cite{prbiru2o7_PRL}}
	\label{iris_cont}
\end{figure}

\begin{figure}
	\includegraphics[width= 8.5cm]{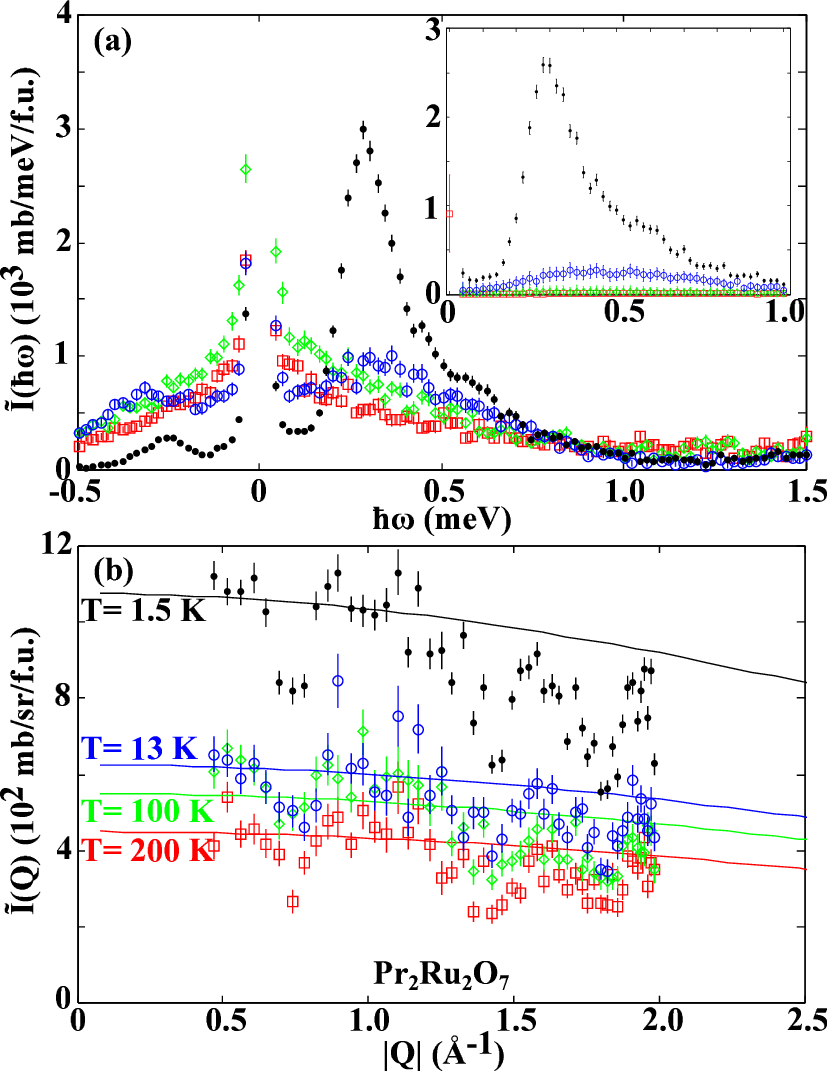}
	\caption{Low energy  Q integrated (a) and $\hbar\omega$ integrated (b) inelastic neutron
	scattering intensity of $x= 0$ at 1.5 K ({$\bullet$}), 13 K
	(\textcolor{blue}{$\circ$}), 100 K (\textcolor{green}{$\diamond$}) and 200 K
	(\textcolor{red}{$\Box$}). Data were obtained by integrating the spectra shown in
	Fig.~\ref{iris_cont} over the range 0.5 \(\leq |Q| \leq\) 1.5 \AA$^{-1}$ and 0.1 \(\leq
	\hbar\omega \leq\) 1.0 meV respectively. For (a) the energy binning is 2 ($\Delta \hbar\omega= 0.02$ meV) compared to Figure~\ref{iris_cont}. The dips observed (at all temperatures) at 0.7 \AA$^{-1}$, $\sim$1.4 \AA$^{-1}$ and $\sim$1.8 \AA$^{-1}$ in (b) are due to variations in detector channel sensitivity (see text for more details). The solid lines shows $|F(Q)|^2$ calculated for Pr$^{3+}$ scaled to the data. Insert shows the same data as in (a) multiplied by the Bose factor.}
	\label{iris_QE}
\end{figure}

To better understand the rare earth magnetism in Pr$_{2-x}$Bi$_{x}$Ru$_2$O$_7$ we carried out low energy inelastic neutron scattering measurements on $\rm Pr_2Ru_2O_7$ using the IRIS spectrometer. Figure~\ref{iris_cont} shows inelastic neutron scattering at 1.5 K, 13 K, 100 K ( T $<$ T$_N$) and 200 K (T $>$ T$_N$). At 1.5 K a sharp mode centered at $\sim 0.25$ meV is observed. At this temperature the magnetic moments on the Ru sublattice are ordered and all the CF excitations are accounted for at higher energies. The absence of any dispersion and indeed of any apparent wave vector dependence to the scattering cross section, beyond that expected from the magnetic form factor of the praseodymium ion (Fig.~\ref{iris_QE}(b)),  shows that this mode is a single ion property. Even a local cluster excitation within the frustrated spin system (zero energy mode) is not viable as that would result in Q-dependent intensity from the cluster structure factor. 


 It is interesting then that the Q-integrated local spectrum is not resolution limited (Fig.~\ref{iris_QE}(a)). At 1.5 K the observed feature appears to consist of two components, a sharp one centered at $\sim$ 0.25 meV and a broad one centered at $\sim$ 0.5 meV. As the temperature is increased the sharp feature appears to decrease in intensity and to collapse into the broad one; increasing the temperature further results in the disappearance of the broad feature. The energy integrated part of the spectrum (Fig.~\ref{iris_QE}(b)) also shows a decrease in intensity with increasing temperature without any apparent change in the dispersion along $Q$. The dips observed (at all temperatures) at 0.7 \AA$^{-1}$, $\sim$1.4 \AA$^{-1}$ and $\sim$1.8 \AA$^{-1}$ coincide with dips in the nuclear incoherent scattering for the same detectors. As a result these (sharp) modulations are extrinsic and may be due to variations in detector channel sensitivity (e.g. shading of certain detectors by the radial collimator, unstable detector electronics etc.) that we were unable to correct for in the data treatment. Even so, the dispersion appears to follow well the squared single ion form factor of Pr$^{3+}$ (shown as solid lines in Fig.~\ref{iris_QE}(b)) at all temperatures, indicating that there are no (or very weak) spatial correlations between the Pr sites in this material.

\begin{figure}
	\includegraphics[width= 8.5cm]{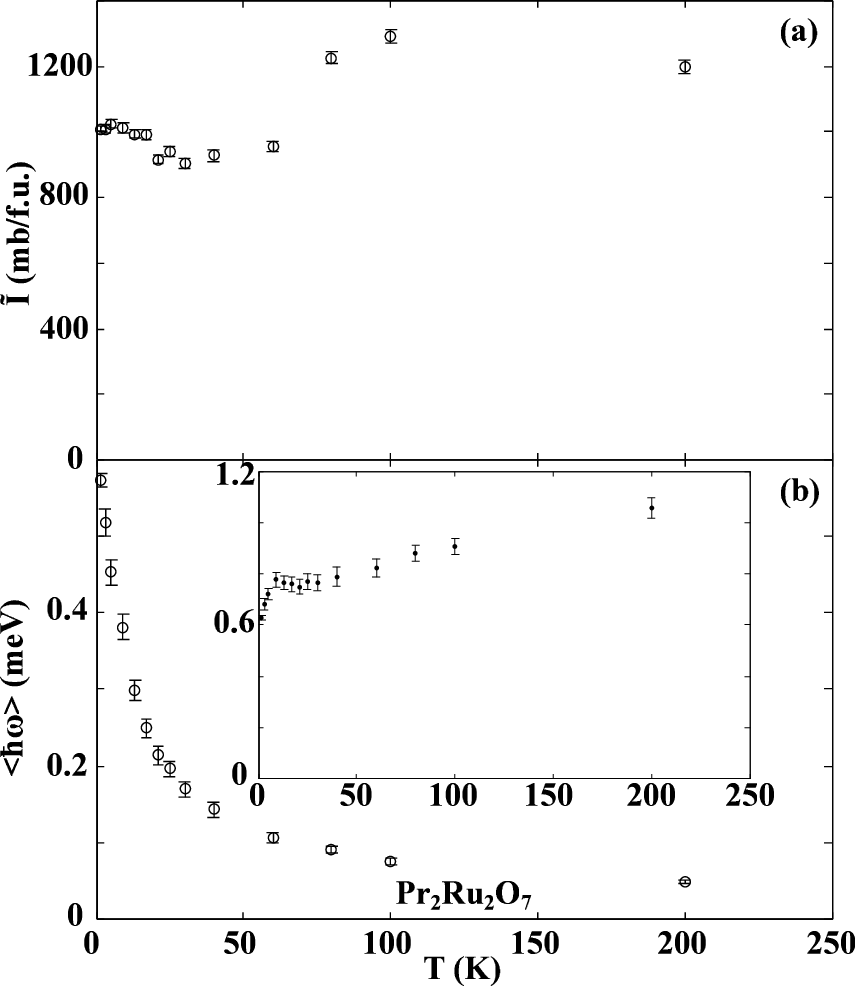}
	\caption{Temperature dependence of the integrated intensity (a) and average energy (b) of the low energy magnetic response  of $x=0$, the insert shows the average energy without detailed balance applied. The region of integration is over 0.5 \(\leq |Q| \leq\) 1.75 \AA$^{-1}$ and 0.1 \(\leq \hbar\omega \leq\) 2.75 meV.}
	\label{sumrule}
\end{figure}

Exact sum rules for $\tilde{I}(Q, \hbar\omega)$can be used to obtain additional "model independent" information temperature dependence of the low energy magnetic response.~\cite{sum_rule} The following rules have been applied to determine the temperature dependence of the integrated intensity and average energy of the low energy magnetic response 
\begin{equation}
\tilde{I}= \int_{-\infty}^{\infty}\tilde{I}(\hbar\omega)\hbar d\omega \approx  \int_{0}^{\infty} (1+e^{-\beta\hbar\omega})\tilde{I}(\hbar\omega)\hbar d\omega
\end{equation}
\begin{eqnarray}
<\hbar\omega>&=& \frac{\int_{-\infty}^{\infty}\hbar\omega \tilde{I}(\hbar\omega) d\omega}
{\int_{-\infty}^{\infty}\tilde{I}(\hbar\omega) d\omega} \nonumber \\
 & \approx& \frac{\int_{0}^{\infty}\hbar\omega (1-e^{-\beta\hbar\omega}) \tilde{I}(\hbar\omega) d\omega}{\int_{0}^{\infty}(1+e^{-\beta\hbar\omega})\tilde{I}(\hbar\omega) d\omega}.
\end{eqnarray}
The result for all measured temperatures is shown in Figure~\ref{sumrule}.  The drop observed in the integrated intensity below 100 K is a result of the integration limits and is due to the disappearance of quasi elastic scattering coming from the ordered moments on the Ru sublattice. There is a gapped excitation of $\sim$ 8 meV associated with the ordered Ru moments (see Fig.~\ref{mr_35meV}b); as the temperature is lowered below this energy, the first excited state becomes depopulated and the quasi-elastic scattering associated with it disappears. The temperature dependence of the average intensity is consistent with what one would expect for a system that has a gapped excitation centred at $\sim$ 0.5 meV.

\section{Analysis and Discussion}

\subsection{EXAFS data}
To obtain more quantitative information on the presence of local structural distortions, we have fitted the low temperature EXAFS data using theoretical functions for each atom-pair, calculated with FEFF8.2.\cite{FEFF8} In such fits the pair-distance and the broadening of the pair distribution function, $\sigma$, for each peak are varied. The coordination number we obtain from the known pyrochlore structure and diffraction results. In addition, the edge energy is varied slightly to correspond to the point on the edge at which the photo-electron wavenumber, k, is zero. The overall amplitude, using the parameter S$_o^2$, was also allowed to vary. This parameter takes into account multiple scattering contributions to the edge height and is typically between 0.7 and 1.0. In our analysis, we obtain an average value for S$_o^2$ from fits to a number of low temperature scans - Pr L$_{\rm III}$; S$_o^2$=0.97, Bi L$_{\rm III}$; S$_o^2$= 1.0, and Ru K; S$_o^2$ = 1.0; however because of the large positive correlation between S$_o^2$ and $\sigma^2$, there is a large systematic error in this parameter. We have used the above values of S$_o^2$ for a given edge for comparison purposes.

\begin{figure}
\includegraphics[width=8.5cm,clip]{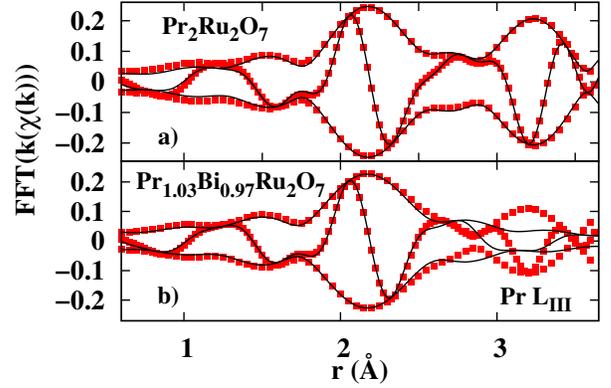}
\caption{a) Fit of the Pr L$_{\rm III}$ edge data at 4 K for $x=0$ from 1.8-3.8
\AA, for a FT range 3.5-10 \AA$^{-1}$.  b) Fits to the Pr-O peak only, for the
mixed sample, $x=0.97$, from 1.6-2.5 \AA. Again the FT range is 3.5-10
\AA$^{-1}$; solid squares are the data while solid lines are the fits.}
\label{pr-fits}
\end{figure}

In most of the following fits we focus primarily on the nearest neighbor metal-O peaks, as they are easiest to fit, except for the pure $x=0$ sample, where we first show a detailed fit out to $\sim$ 4 \AA{} for the Pr L$_{\rm III}$ edge. This fit includes Pr-O1, Pr-O2, the first metal-metal pairs (Pr-Pr and Pr-Ru), longer Pr-O pairs, and multi-scattering peaks (see Fig.~\ref{pr-fits}a). We constrained the distances to be consistent with the crystal structure (allowing for an overall expansion of the unit cell) and used the coordination numbers from the structure to reduce the errors in $\sigma$. We obtain a very good fit out to 3.8 \AA{} that is consistent with diffraction results; see Table~\ref{pr_fit}, for the parameters of the first two O shells for comparison with fits of the mixed sample. For more distant peaks we would need to add additional neighbors to account for longer pair distances; for example, the fit at 4 \AA{} is poor because longer Pr-O, Pr-Pr, Pr-Ru, and multi scattering peaks are not included.

For the mixed sample, we only fit the Pr-O peak as shown in Figure~\ref{pr-fits}b. The amplitude of this peak is comparable to that for the pure $x=0$ sample - the widths change slightly (See Table~\ref{pr_fit}). This shows that the Pr-O1 and Pr-O2 bond lengths are only slightly distorted in the mixed sample. However, note the much smaller amplitude for the second main peak (a sum of Pr-Pr, Pr-Ru, and Pr-Bi) for the mixed sample near 3.2 \AA.

\begin{figure}
\includegraphics[width=8.5cm,clip]{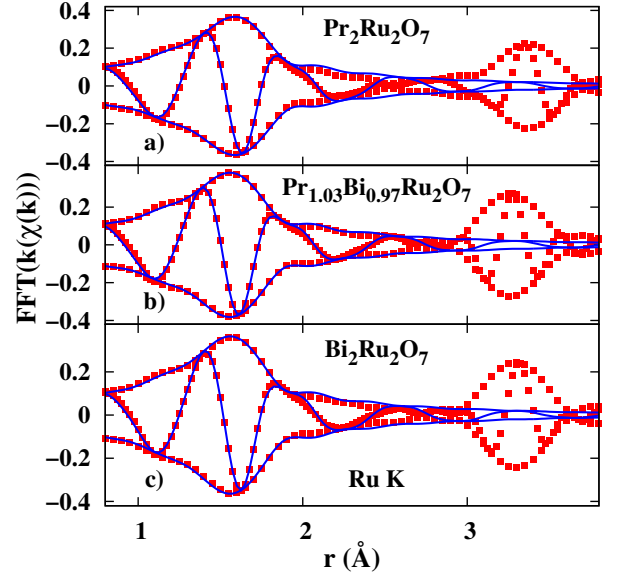}
\caption{Fits of the Ru-O1 peak for a) $x=0$, b) $x=0.97$, and c) $x=2$; data square points; fit of first Ru-O peak - solid lines. The Ru-O peaks have nearly the same amplitude indicating very little disorder of Ru-O in any sample.  FT range 4.5-14 \AA$^{-1}$; fit range 1.3-2 \AA.}\label{Ru-edge-fit}
\end{figure}

The Ru K-edge data show  similar behavior to the Pr L$_{\rm III}$ edge data. First a good fit out to 4 \AA (not shown) can be obtained, again showing a well ordered structure. In Figure~\ref{Ru-edge-fit}a-c we compare the fits for the Ru-O1 peak in the Ru data collected for $x=0$, $x=0.97$, and $x=2$. The fit range used was 1.3-2 \AA, but a good fit extends below 1 \AA.  Above $\sim$2 \AA, the tails of higher peaks partially interfere destructively with the Ru-O peak, but the agreement is still quite good. In each case the Ru-O1 peak is large, indicating a well ordered structure. As for the Pr L$_{\rm III}$ edge results, the Ru-O1 bond length agrees with diffraction to better than 0.01 \AA{} and are not tabulated; the values of $\sigma$ at 4K are given in Table \ref{ru_fit}, and are identical within our errors. However the further neighbor peak near 3.3 \AA{} changes, the amplitude grows as the Bi concentration increases, most likely as a result of a change in interference. For $x=0$, the Ru-Ru and Ru-Pr peaks are partially out of phase leading to a reduced amplitude; in $x=0.97$, the Ru-Pr amplitude is reduced by $\sim$ 50\% (and the Ru-Bi is disordered), and hence there is less destructive interference. Because there is a changing mixture of Ru-Ru, Ru-Pr and Ru-Bi pairs with increasing Bi concentrations, a more quantitative characterization of the disorder in the metal-metal peaks requires a more detailed fit, which is beyond the scope of this paper.

\begin{table*}

\caption{Results for the A-site: Pr-O and Bi-O peaks from fits for the Pr and
Bi L$_{\rm III}$ edges of $x=0$, $x=0.97$ and $x=2$. The fit ranges for the Pr
L$_{\rm III}$ edge are: $x=0$, 1.8 - 3.8 \AA, and $x=0.97$, 1.6-2.5 \AA{} (Pr-O
peak only). The ranges for the Bi fits (O1 and O2 shells only) are: $x=2$,
1.2-2.8 \AA{} and $x=0.97$, 1-2.4 \AA. Estimated errors on r, $\pm$ 0.01 \AA;
estimated errors for $\sigma^2$, including systematic errors which dominate,
$\pm$ 0.0005 \AA$^2$. The diffraction results in last column are from this work
for $x=0$ and from Avdeev {\it et al.}~\cite{Avdeev02}, (model h with an
average position for O2) for $x=2$. The Bi offcenter displacement, D, for $x=2$
is 0.16 \AA{} which is identical to the diffraction results of Avdeev {\it et
al.}~\cite{Avdeev02} within our errors and also agrees with Shoemaker {\it et
al.}~\cite{Shoemaker11}; for $x=0.97$, D = 0.17 $\pm$ 0.02 \AA.}

\begin{tabular}{|c|c|c|c|c|}
\hline
 Compound & Atom Pair & $\sigma^2$ (\AA$^2$) &  r (EXAFS) (\AA)  &  r (diffraction) (\AA)\\
        \hline
 $x=0$ & Pr-O2 & 0.0058 & 2.27 & 2.25  \\
 $x=0$ & Pr-O1 & 0.0033 & 2.56 & 2.54  \\
 $x=0.97$ & Pr-O2 & 0.0049 & 2.23 & 2.23  \\
 $x=0.97$ & Pr-O1 & 0.0031 & 2.54 & 2.54  \\
 $x=0.97$ & Bi-O2 & 0.0031 & 2.258 & - \\
 $x=0.97$ & Bi-O1a & 0.015 & 2.445 & - \\
 $x=0.97$ & Bi-O1b & 0.0048 & 2.596 & - \\
 $x=0.97$ & Bi-O1c & 0.013 & 2.738 & - \\
 $x=2$ & Bi-O2 & 0.0046 & 2.29 & 2.234 \\
 $x=2$ & Bi-O1a & 0.0020  & 2.409 & 2.410 \\
 $x=2$ & Bi-O1b & 0.0016 & 2.552 & 2.554 \\
 $x=2$ & Bi-O1c & 0.0056 & 2.686 & 2.690 \\
        \hline
\end{tabular}
\label{pr_fit}
\end{table*}

\begin{table}
\caption{Results from fits of the Ru-O1 peak in Ru K-edge data for for $x=0$, $x=0.97$, and $x=2$;
fit range 1.3-2 \AA. Relative errors in $\sigma^2$ $\pm$ 0.0004 \AA$^2$.} 
\begin{tabular}{|c|c|c|}
\hline
Compound & Atom Pair & $\sigma^2$ (\AA$^2$)\\
       \hline
$x=0$ & Ru-O1 & 0.0027 \\
$x=0.97$ & Ru-O1 & 0.0028  \\
$x=2$ & Ru-O1 & 0.0026 \\
       \hline
\end{tabular}
\label{ru_fit}
\end{table} 

Comparing the mean-squared atomic displacements obtained from the fits to the neutron
diffraction data  (Table~\ref{tab:nucl_fit}) with the $\sigma^2$ values obtained
for Pr-O1, Pr-O2 and Ru-O1 (0.0033 \AA$^2$, 0.0058 \AA$^2$ and 0.0027 \AA$^2$
respectively) obtained from our EXAFS analysis (Tables~\ref{pr_fit}
and~\ref{ru_fit}) for $x=0$ we can observe the following. The disorder in
the Pr-O1 and Ru-O1 bonds is very small while that for Pr-O2 is about twice as
large, also the ratio of the $<u^2>$ parameters for Pr and the O1/O2 atoms is close
to 2.  This indicates that the O1 atoms have little disorder. The two large
quantities are $<u^2>$ for Pr and  $\sigma^2$ for Pr-O2. If Pr is displaced a
little along the Pr-O2 axis it will only affect the Pr-O2 and Pr$<u^2>$
parameters. This suggests that there is some intrinsic disorder on the Pr site
in the pyrochlore structure which could influence the ground state
properties. 

\begin{figure}
\includegraphics[width=8.5cm,clip]{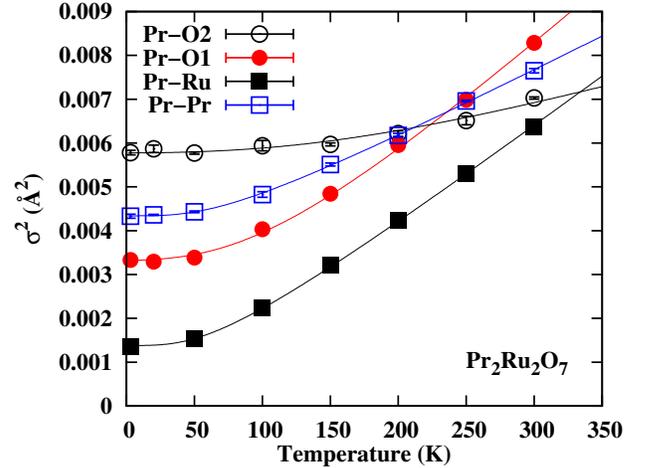}
\caption{Plots of $\sigma^2$ vs T for the Pr-O1, Pr-O2, Pr-Ru, and Pr-Pr pairs
from EXAFS data at the Pr L$_{\rm III}$ edge of $x= 0$. The Pr-O2 pair has a significant
static contribution to $\sigma^2$ at low T ($\sim$ 0.0033 \AA$^2$) compared to
the Pr-O1 pair. A similar behavior is observed when comparing the Pr-Ru and
Pr-Pr pairs; there is a significant static contribution to $\sigma^2$ for the
Pr-Pr pair ($\sim$ 0.0027 \AA$^2$) but not the Pr-Ru pair at 4 K. The solid
lines are fits to a correlated Debye model. The correlated Debye temperatures
are: Pr-O1 - 520(30) K; Pr-O2 - 880(50) K: Pr-Ru - 290(5) K; Pr-Pr - 316(5) K. Relative errors are indicated by (); absolute errors, mainly from systematic effects are $\sim$ 10 \%.}
\label{pr-sigma}
\end{figure}

To further explore possible disorder on the Pr site we carried out a
temperature dependent EXAFS study at the Pr L$_{\rm III}$ edge. The temperature
dependence of the Debye-Waller factor $\sigma^2$(T) provides an estimate of the
zero-point motion contribution to $\sigma^2$ at low T; if there is significant
static disorder, $\sigma^2$(4 K) will be larger than expected.  In
Fig.~\ref{pr-sigma} we plot $\sigma^2$(T) for the Pr-O1, Pr-O2, Pr-Ru, and
Pr-Pr pairs. The solid lines are fits to a correlated Debye model. The Pr-O2
pair is a much stiffer bond (low slope) but has a significant static
contribution to $\sigma^2$ at low T $\sim$ 0.0033 \AA$^2$. A similar behavior is observed
for the Pr-Ru and Pr-Pr pairs which have the same pair distance. The Pr-Ru PDF
has little static disorder while $\sigma^2$ for the Pr-Pr pair has a large
static contribution at 4K. The lack of significant disorder for Pr-O1, Pr-Ru,
and earlier Ru-O1, suggests that the disorder is primarily along the Pr-O2
axis. Since $<u^2>$ for O1 and O2 are comparable and much smaller than $<u^2>$
for Pr (Table~\ref{tab:nucl_fit}), most of the disorder must be about the Pr
site. Assuming a Pr displacement along the Pr-O2 axis, the magnitude is $\sim$
0.05 - 0.06 \AA. 

The data and simulations presented in Figures~\ref{r-space-pure}b and~\ref{r-space-mixed}b show considerable disorder of the Bi-O1 shell. Since the Pr-O1 and Ru-O1 pair distributions are ordered (See Fig.~\ref{r-space-pure}a, c), this indicates that the disorder for Bi-O1 arises from displacements of Bi from the ordered A-site position, either away from or towards the ring of O1 atoms, i.e. in a direction perpendicular to the Bi-O2 axis in the Bi$_4$O$_2$ tetrahedra. There may also be small, correlated, translation-rotations of the A-tetrahedra containing the Bi which could be accommodated by changes in the Pr-O1-Ru angles with little disorder of the Ru-O1 and Pr-O1 bonds, consistent with the Pr and Ru EXAFS discussed above.

Diffraction studies find a displacement of the Bi away from the Bi-O2 axis in the end compound $x=2$, and attributed it to the 6s lone pair electrons on Bi$^{3+}$.~\cite{Avdeev02} The distortion has been modeled by allowing the Bi to move off-center in six equivalent directions and then set the filling fraction at 1/6. For example one direction for Bi to move off-center is towards a Ru atom, or midway between two O1 atoms (six possibilities) - this is called the h-model, and the site is 96h (0,y,-y) in space group $Fd\bar{3}m$. They also considered a similar model with the six off-center directions rotated by $\sim$30$^\circ$ i.e. approximately displaced towards the midpoint between two Ru atoms or roughly towards an O1 atom - this is called the g-model; site 96g (x,x,z). For this model, the off-center displacements are not quite perpendicular to the undistorted Bi-O2 axis and the ring of displaced sites is slightly corrugated. In our first fits we tried just a broad distribution for Bi-O1. These do not fit well and discrete Bi-O1 distances are required as indicated in the diffraction studies.

The diffraction results also suggest that the O2 atoms are displaced along four symmetry directions (with occupancy 1/4 for each off-center site).~\cite{Avdeev02} Assuming that the Bi and O2 off-center displacements are uncorrelated, this leads to a very broad distribution of Bi-O2. We have tried this distribution for O2 and it does not fit our EXAFS data; although there is some broadening of the Bi-O2 distribution it is much smaller than suggested from uncorrelated displacements of Bi and O2. In the models we compare below we use a single peak for Bi-O2 but allow it to broaden slightly.
\begin{figure}
\includegraphics[width=8.5cm,clip]{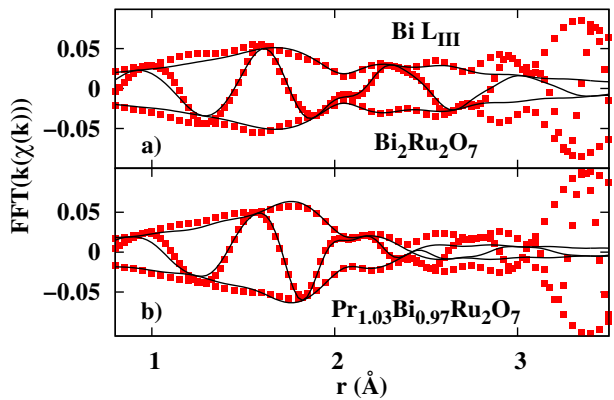}
\caption{Fits of the Bi-O peak (Bi-O1 and Bi-O2) for a) $x=2$ and b) $x=0.97$, using a single peak for Bi-O2 and the split 222 or h model for Bi-O1. The three peaks in the 222 model are split by $\delta r$ $\sim$ 0.16 \AA. For the mixed compound the Bi-O2 peak which is located near 1.8 \AA, has slightly less broadening than in the pure Bi end compound, but the disorder of the individual peaks for Bi-O1 is larger. Fit range 1.2-2.8 \AA{} for $x=2$; 1.0-2.4 \AA{} for $0.97$.}
\label{Bi-edge-fit}
\end{figure}

For the h-model there are three Bi-O1 peaks at r$_0$ and r$_0$ $\pm \delta r$, each with two O1 neighbors - thus the numbers of neighbors in the peaks are in the ratio 2:2:2 and we refer to it as the 222 model.  For the g-model there are four Bi-O1 distances, one O1 neighbor each at r$_0$ $\pm \delta_1 r$ and two neighbors each at r$_0$ $\pm \delta_2 r$; we therefore call this model the 1221 model (the ratio of the O1 coordinations). In fitting the O-peak one needs to remember that when there are two quite close bond lengths (here the Bi-O1 and Bi-O2, or Pr-O1 and Pr-O2) then there will be interference between the two components in $r$-space. The dip at 1.7-1.8 \AA{} for the Pr data and near 2.1 \AA{} for the Bi data are the results of this interference.

We have carried out fits using both the 222 and 1221 models above (h and g models in diffraction). The fits were similar, with the h-model slightly better; however the improvement in the goodness of fit parameter was not statistically better based on the Hamilton F-test;~\cite{Downward07} thus Bi-O1 can be quite well modeled using either distribution. However Shoemaker {\it et al.}~\cite{Shoemaker11}  find the h-model is better from nuclear density plots for $x=2$. Consequently we only show results for this model. In Figure~\ref{Bi-edge-fit}a we show the fits of the Bi-O peak for $x=2$ and in Figure~\ref{Bi-edge-fit}b the fit for the mixed compound $x=0.97$. The fit ranges are 1.2-2.8 and 1.0-2.4 \AA{} respectively. The data and fits show that for the short Bi-O2 peak, the pure compound is slightly more disordered. In contrast for the Bi-O1 peak, the amplitude from 2-2.8 \AA{} is lower for the mixed compound indicating more disorder in this material. The fits also have a larger broadening of the three individual split Bi-O1 peaks. Surprisingly the splitting for the mixed sample is about the same - 0.17 \AA{} within our uncertainty, $\pm$ 0.02 \AA. Some parameters are provided in Table~\ref{pr_fit}.

\subsection{Single ion properties}
\begin{figure}
\includegraphics[width= 8.5cm]{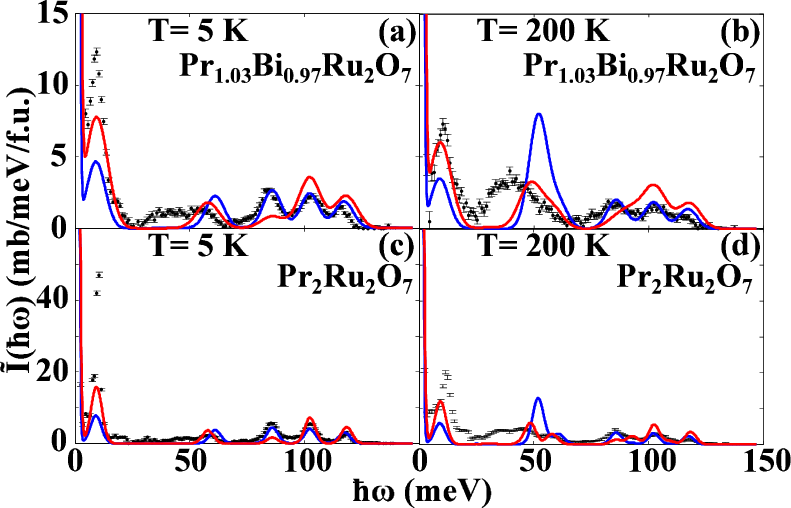}
\caption{Combined $E_i$= 35 meV and 160 meV magnetic response of Pr$_{2-x}$Bi$_{x}$Ru$_2$O$_7$ for $x= 0.97$ and $x= 0$ at 5 K (a, c) and 200 K (b,d)($\bullet$). The solid lines show the calculated spectra for Model 1 (blue line) and Model 2 (red line) using the fitted CF parameters listed in Table~\ref{tab:pbro_cf_par}, including an intrinsic Gaussian broadening of the transitions. Only this intrinsic Gaussian broadening has been allowed to vary between $x= 0.97$ and $x= 0$.\label{pbro_cf_models}}
\end{figure}

In order to explain the observed local low energy spin excitations in Pr$_{2-x}$Bi$_{x}$Ru$_2$O$_7$ near the metal to insulator transition, as well as the metallic spin-liquid behavior in Pr$_2$Ir$_2$O$_7$, it was important to determine the relevant low energy spin degrees of freedom in these systems.~\cite{prbiru2o7_PRL, pr2ir2o7_PRL} For this we have analyzed the high energy magnetic response of $x= 0.97$, $x= 0$ and Pr$_2$Ir$_2$O$_7$ (Figs.~\ref{mr_35meV} and~\ref{mr_160meV}) to determine the ground-state and CF levels of Pr in these materials. Preliminary analysis has already shown the presence of 5 CF excitations, consistent with Pr$^{3+}$ being in a CF of symmetry $D_{3d}$ whose CF Hamiltonian is given by equation~\ref{eq:CF_Ham}. For such a simple CF Hamiltonian the dynamic spin correlation function given in equation~\ref{eq:dyn_spin_cor} for a transition from the CF state $|p>$ to $|q>$ can be simply rewritten to be
\begin{eqnarray}
S^{\alpha \alpha}({\bf Q}, \omega)&=&
\sum_{p,q} \frac{2}{3} \rho_p |<p|J^{\alpha}|q>|^2 \nonumber \\
 & & \times \delta(E_p-E_q + \hbar\omega)
\label{eq:nc_cf}
\end{eqnarray}
where $\rho_p$ is the occupancy of the state $|p>$ with energy $E_p$. A Monte-Carlo search of the CF parameter space was performed to obtain an initial set of CF parameters used to fit the data. For the Monte-Carlo search and the fitting of the CF parameters the spectra with incident energies $E_i$= 35 meV and 160 meV were combined into one spectra (Figs.~\ref{pbro_cf_models} and~\ref{pio_cf_models}). Problems arose during the analysis of the CF excitations due to the additional broadening of the CF level excitation at around 50 meV compared to the other CF level excitations, which cannot be accounted for by the simple single-ion CF Hamiltonian given in equation~\ref{eq:CF_Ham}. One possible explanation for such a broadening is magneto-elastic coupling.

In general phonons and CF transitions are considered to be decoupled and as such the measured spectra of both phenomena can be determined and interpreted independent of each other. However, if the energy separation of the levels within the ground-state multiplet are comparable with the energies of strong phonon modes, coupling of the two systems may occur. Such a CF-phonon coupling, observed by neutron scattering, is already known from CeCu$_2$ , YbPO$_4$ and CeCuAl$_3$.~\cite{CeCu2_1, CeCu2_2,YbPO4_1,YbPO4_2, CeCuAl3} It can result in a broadening and over damping of energy levels that cannot be explained by the simple single-ion CF model. From Figure~\ref{phsub} it can be seen that the 50 meV CF excitation lies on top of a large phonon background which is still clearly visible at low scattering angles, while the other CF excitations do not. It is not unreasonable to assume that the broadening which is observed in these two systems for the 50 meV CF excitation is due to CF-phonon coupling. Additional (single crystal inelastic neutron scattering and Raman scattering) experiments are needed to determine whether or not this assumption is correct. In the current analysis of the high energy magnetic response the possibility of CF-phonon coupling has not been taken into account and only the simple single-ion CF Hamiltonian (eq.~\ref{eq:CF_Ham}) has been used. Such a simple single-ion CF model also does not account for the additional broadening that is observed for all CF level excitations in Pr$_{2-x}$Bi$_{x}$Ru$_2$O$_7$ for $x= 0.97$ compared to $x= 0$.

\subsubsection{Pr$_{2-x}$Bi$_x$Ru$_2$O$_7$}
\begin{table}
\caption{Fitted CF parameters of Pr$_{2-x}$Bi$_{x}$Ru$_2$O$_7$ for $x= 0$. The parameters were obtained from fits to the magnetic response at 5 K. All parameters are in meV.\label{tab:pbro_cf_par}}
\begin{tabular}{ccc} \hline \hline
 & Model 1 & Model 2  \\ \hline
$B^0_2$ & -8(1)$\times10^{-1}$  & -1.3(2) \\
$B^0_4$ & -4.2(5)$\times10^{-2}$ & -2(1)$\times10^{-3}$ \\
$B^3_4$ & 2.9(3)$\times10^{-1}$ & 6.4(8)$\times10^{-1}$ \\
$B^0_6$ & 7.7(2)$\times10^{-4}$ & 8.3(3)$\times10^{-4}$ \\
$B^3_6$ & 3(3)$\times10^{-3}$ & 1.09(6)$\times10^{-2}$ \\
$B^6_6$ & 4.1(8)$\times10^{-3}$ & 6(1)$\times10^{-3}$ \\ \hline
\end{tabular}
\end{table}
Two sets of CF parameters were found from the Monte-Carlo search of parameter space to give excitations at the observed energies. These two sets of CF parameters have been fitted to the 5 K spectrum of the pure material, allowing for an intrinsic Gaussian broadening of the CF transitions. Both models gave similar fits to the data. The refined values of the individual CF parameters for both models are listed in Table~\ref{tab:pbro_cf_par}, while the corresponding energy levels and eigenvectors are given in Table~\ref{tab:pbro_cf_eigenvect}. It can be seen from Table~\ref{tab:pbro_cf_eigenvect} that these two models give a slightly different level scheme for the excitations. Model 1 has a doublet ground-state, followed by a singlet, a doublet, 2 singlets and a doublet, while Model 2 has a doublet ground-state, followed by 3 singlets and 2 doublets. Interestingly both models give a doublet ground-state and a singlet first excited state. The symmetry of these two states is the same in both models. The suggestion that the Pr ions have a doublet ground-state is confirmed by the observation that it is split due to either a low density of
extended defects or a density wave which generates a continuum of local environments in the doped material.~\cite{prbiru2o7_PRL}

\begin{table}
\caption{Energies (E$_i$ in meV) and CF wave functions ($\psi_i$) of the 9-fold degenerate ground-state multiplet $^4H_3$ of Pr$_{2-x}$Bi$_{x}$Ru$_2$O$_7$. The CF level energies and wave functions were calculated for both models using the CF parameters  listed in Table~\ref{tab:pbro_cf_par}.($<$) represents a CF doublet level.\label{tab:pbro_cf_eigenvect}}
\begin{tabular}{cc} \hline \hline
E$_i$ & $\Psi_i$ \\ \hline
\multicolumn{2}{c}{Model 1}\\
0$<$ & \(\psi_g= 0.935|\mp4> - 0.073|\pm2> \pm 0.348|\mp1>\) \\
9.10 & \(\psi_1= -0.166|3> + 0.972|0> + 0.166|-3>\) \\
61.02$<$ & \(\psi_2= \mp0.348|\mp4> \mp0.013|\pm2> + 0.937|\mp1>\) \\
86.02 & \(\psi_3= 0.686|3> + 0.235|0> - 0.687|-3>\) \\
102.41 & \(\psi_4= 0.707|3> + 0.707|-3>\) \\
117.44$<$ & \(\psi_5= 0.063|\mp4> + 0.989\pm2>\pm0.039|\mp1>\) \\ \hline
\multicolumn{2}{c}{Model 2} \\
0$<$ & \(\psi_g= 0.860|\mp4> - 0.121\pm2> \pm0.495|\mp1>\) \\
9.32 & \(\psi_1=-0.626|3> + 0.465|0> + 0.626|-3>\)  \\
57.92 &  \(\psi_2= 0.707|3> + 0.707|-3>\) \\
86.15 & \(\psi_3= 0.329|3> + 0.885|0> -0.329|-3>\) \\
102.49$<$ &  \(\psi_4= 0.017\mp4> + 0.978|\pm2> \pm0.210|\mp1>\) \\
118.47$<$ & \(\psi_5= \pm0.509|\pm4> \pm0.173|\mp2> + 0.843|\pm1>\) \\ \hline
\end{tabular}
\end{table}

Figure~\ref{pbro_cf_models} shows the calculated spectra of both models, compared with the magnetic response of the pure and dilute material at 5 K and 200 K. For the dilute material only the intrinsic Gaussian broadening has been allowed to vary. At 5 K the possibility of an internal magnetic field due to the ordering of the Ru sublattice in the pure material has not been taken into account in the calculated spectra. It can be seen from this Figure that, even though the CF level energies that have been obtained are close to those observed in the magnetic response of both materials, there is a large discrepancy between the observed and calculated spectra. At both 5 and 200 K Model 1 gives a better description of the three excitations around 100 meV energy transfer then Model 2, while Model 2 gives a slightly better description of the excitation at 10 meV energy transfer. As expected both models have problems describing the broad excitation centred at around 50 meV energy transfer. They do however have similar temperature dependence as is observed for the measured magnetic response, i.e. both allow for a transition from the first excited state at 10 meV to the second excited state at around 50 meV energy transfer. 

\subsubsection{Pr$_2$Ir$_2$O$_7$}
\begin{figure}
\includegraphics[width= 8.5cm]{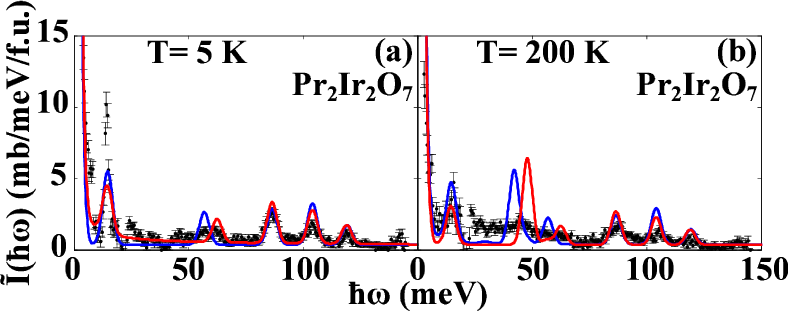}
\caption{Combined $E_i$= 35 meV and 160 meV magnetic response of Pr$2$Ir$_2$O$_7$ at 5 K (a) and 200 K (b)($\bullet$). The solid lines show the calculated spectra for Model 1 (blue line) and Model 2 (red line) using the fitted CF parameters listed in Table~\ref{tab:pio_cf_par}, including an intrinsic Gaussian broadening of the transitions.\label{pio_cf_models}}
\end{figure}
Due to the strong absorbing nature of the Pr$_2$Ir$_2$O$_7$ sample it was found that a  scaling factor was needed to combine the $E_i$= 35 and 160 meV spectra. This scaling factor was obtained by integrating the elastic line in a region ($1.4 \leq Q \leq 1.625$ \AA) on the low angle bank where the spectra overlap and determined to be $\tilde{I}(E_i= 160 ~\mathrm{meV})/\tilde{I}(E_i= 35 ~\mathrm{meV})= 1.27$. As was the case for Pr$_{2-x}$Bi$_x$Ru$_2$O$_7$ two sets of CF parameters were found from the Monte-Carlo search of parameter space to give excitations at the observed energies. These two sets of CF parameters have been fitted to the 5 K spectrum, allowing for an intrinsic Gaussian broadening of the CF transitions. The refined values of the individual CF parameters for both models are listed in Table~\ref{tab:pio_cf_par}, while the corresponding energy levels and eigenvectors are given in Table~\ref{tab:pio_cf_eigenvect}. It can been seen from Table~\ref{tab:pio_cf_eigenvect} that these two models are very similar and only differ in the assignment of the two highest energy excitations. Both models give a doublet ground-state, followed by a singlet, a doublet and a singlet in Model 1 this is then followed by a doublet and a singlet, while for Model 2 this is reversed. The symmetry of the doublet ground-state and the first three excited states is the same in both models.
\begin{table}
\caption{Fitted CF parameters of Pr$_2$Ir$_2$O$_7$. The parameters were obtained from fits to the magnetic response at 5 K. All parameters are in meV.\label{tab:pio_cf_par}}
\begin{tabular}{ccc} \hline \hline
 & Model 1 & Model 2  \\ \hline
$B^0_2$ & -6(2)$\times10^{-1}$  & -0.9(1) \\
$B^0_4$ & -4.0(4)$\times10^{-2}$ & -4.2(1)$\times10^{-2}$ \\
$B^3_4$ & -3.0(4)$\times10^{-1}$ & 2.4(6)$\times10^{-1}$ \\
$B^0_6$ & 3.3(6)$\times10^{-4}$ & 7.3(4)$\times10^{-4}$ \\
$B^3_6$ & -0(7)$\times10^{-3}$ & 3(5)$\times10^{-3}$ \\
$B^6_6$ & 8(2)$\times10^{-3}$ & 4(1)$\times10^{-3}$ \\ \hline
\end{tabular}
\end{table}

Figure~\ref{pio_cf_models} shows the calculated spectra of both models compared with the magnetic response of Pr$_2$Ir$_2$O$_7$ at 5 K and 200 K. It can be clearly seen that, even though the CF level energies that have been obtained are close to those observed in the magnetic response, there is a large discrepancy between the observed and calculated spectra. Both models give similar descriptions of the data and mainly differ in the position the excitation centred at around 50 meV. This is likely due to the broad nature of this excitation. Not surprising their temperature dependence is almost identical and similar to what is observed for the measured magnetic response. Both allow for a transition from the first excited state at 15 meV to the second excited state at around 50 meV energy transfer.
\begin{table}
\caption{Energies (E$_i$ in meV) and CF wave functions ($\psi_i$) of the 9-fold degenerate ground-state multiplet $^4H_3$ of Pr$_2$Ir$_2$O$_7$. The CF level energies and wave functions were calculated for both models using the CF parameters  listed in Table~\ref{tab:pio_cf_par}.($<$) represents a CF doublet level.\label{tab:pio_cf_eigenvect}}
\begin{tabular}{cc} \hline \hline
E$_i$ & $\Psi_i$ \\ \hline
\multicolumn{2}{c}{Model 1}\\
0$<$ & \(\psi_g= 0.929|\mp4> - 0.160|\pm2> \mp 0.348|\mp1>\) \\
14.77 & \(\psi_1= 0.233|3> + 0.944|0> - 0.232|-3>\) \\
57.01$<$ & \(\psi_2= \pm0.345|\mp4> + 0.938|\mp1> \mp 0.014|\pm1>\) \\
  &   \( \pm 0.005|\pm2> + 0.005|\pm4>\) \\
86.67 & \(\psi_3= -0.668|3> + 0.329|0> + 0.668|-3>\) \\
104.29$<$ & \(\psi_4= 0.139|\mp4> \pm0.119|\mp2> \mp0.058|\mp1> \) \\
  &  \(+ 0.007|\pm1> + 0.980|\pm2> \pm0.017|\pm4>\) \\
119.41 & \(\psi_5= 0.707|3> + 0.707|-3>\) \\ \hline
\multicolumn{2}{c}{Model 2} \\
0$<$ & \(\psi_g= 0.950|\mp4> - 0.067\pm2> \pm0.306|\mp1>\) \\
14.62 & \(\psi_1= -0.136|3> + 0.981|0> + 0.136|-3>\)  \\
62.61$<$ &  \(\psi_2= \mp0.305|\mp4> + 0.952|\mp1> \pm 0.006|\pm1>\) \\
  &  \(\pm 0.002|\pm2> + 0.002|\pm4>\) \\
86.63 & \(\psi_3= 0.694|3> + 0.192|0> -0.693|-3>\) \\
104.28 &  \(\psi_4= 0.707|3> + 0.707|-3>\) \\
119.17$<$ & \(\psi_5= 0.064|\mp4> \mp0.245|\mp2> \pm0.018|\mp1>\) \\ 
  & \(+ 0.002|\pm1> + 0.990|\pm2> \mp0.016|\pm4>\) \\ \hline
\end{tabular}
\end{table}

While it is clear from Figures~\ref{pbro_cf_models} and~\ref{pio_cf_models} that the simple single-ion CF model does not allow for an adequate description of the data and that there are additional interactions of importance in this system that influence the magnetic response, this analysis does however confirm the Pr ions have a doublet ground-state in both systems. CF-phonon interactions might explain the unequal broadening of the 50 meV excitation, the additional broadening of the excitations in the $x= 0.97$ material could be due to alloying effects. Single crystal experiments are needed to identify these additional interactions and determine how they might couple to the CF level excitations.

\subsection{Magnetic ordering}
\begin{figure}
	\includegraphics[width= 8.5cm]{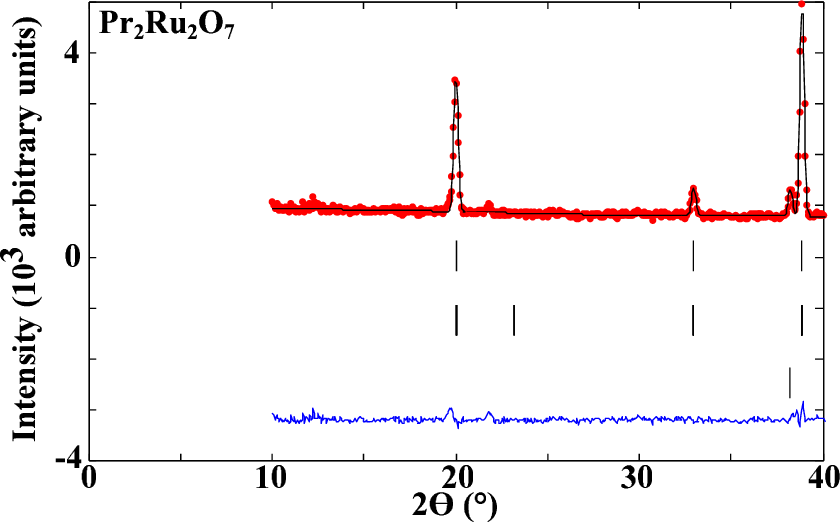}
	\caption{Rietveld fit (solid black line) of the Y$_2$Ru$_2$O$_7$ model to the $x= 0$ neutron powder diffraction profile measured at 1.5 K ($\bullet$). The residual of the fit (blue line) is shown at the bottom of the plot (R$_p$= 10.7\%, R$_{wp}$= 8.63\%, R$_{magn}$= 5.49\%, $\chi^2$= 1.91, $\mu$(Ru)= 1.48(4) $\mu_B$).The upper, middle and lower tick marks indicate Bragg reflections coming from the crystal, magnetic structure of the $x= 0$ and RuO$_2$ impurity phase respectively.}
	\label{d_y2ru27_fit}
\end{figure}

As was noted in section~\ref{sec_diffraction} when cooling the $x= 0$ sample through T$_N$ additional Bragg intensity is observed that cannot be accounted for by nuclear contributions only and must come from long-range ordering of the moments associated with the Ru sublattice. The magnetic Bragg peaks sit on top of nuclear Bragg peaks and can be indexed using a \textbf{k}= (0,0,0) propagation vector and their increase below T$_N$ is similar to that observed for Y$_2$Ru$_2$O$_7$.~\cite{Y2Ru2O7} Indeed, using the model proposed for Y$_2$Ru$_2$O$_7$ gives a very good fit to the data (Fig.~\ref{d_y2ru27_fit}). The ordered Ru moment (1.48(4) $\mu_B$) found is similar to that obtained for Y$_2$Ru$_3$O$_7$ (1.36 $\mu_B$). As the magnetic transition is second order we have performed representational analysis using the SARAh program to gain a more detailed understanding of the Ru ordering in $x= 0$.~\cite{SARAh}

\subsubsection{Representational analysis}
\begin{figure}
	\includegraphics[width= 8.5cm]{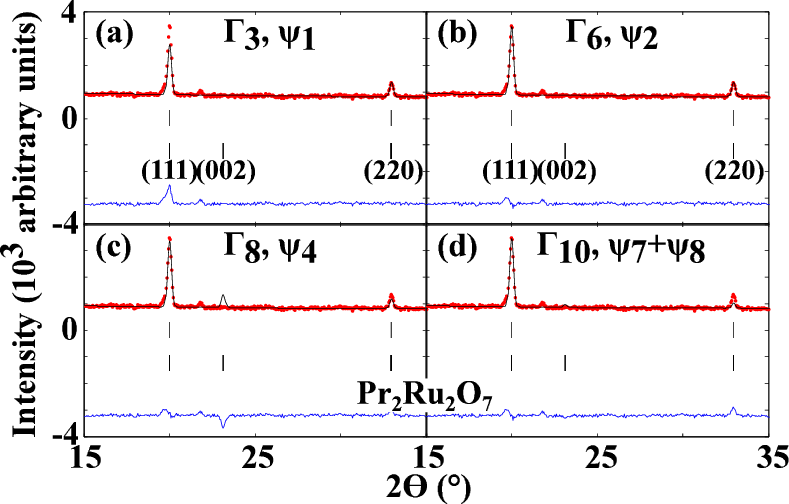}
	\caption{Rietveld fits (solid black lines) of $\Gamma_3$ (a), $\Gamma_6$ (b), $\Gamma_8$ (c) and $\Gamma_{10}$ (d), with ordered moments only on the Ru sublattice, to the $x= 0$ neutron powder diffraction profile measured at 1.5 K ($\bullet$). The residual of the fits (blue lines) is shown at the bottom of the plots. The upper and lower tick marks indicate Bragg reflections coming from the crystal and magnetic structure of $x= 0$ respectively.}
	\label{irrep_fits}
\end{figure}

Representational analysis shows that for space group $Fd\bar{3}m$ with propagation vector \textbf{k}= (0,0,0) the magnetic representation of the Ru (and Pr) sublattice can be decomposed in terms of the following irreducible representations (IRs):
\begin{equation}
	\Gamma_{mag}= \Gamma_3^1 + \Gamma_6^2 + \Gamma_8^3 + 2\Gamma_{10}^3.\label{eq:irreps}
\end{equation}
The corresponding basis vectors (BVs) are listed in Table~\ref{tab:BV}. We have fitted each IR to the diffraction profiles collected at 1.5 K and 60 K. This was done to look for evidence of possible Pr ordering at low temperatures. It was found for all IRs that adding an ordered moment on the Pr sublattice did not significantly improve the fit to the data and that the ordered Pr moment (up to $\sim$0.3 $\mu_B$) is much lower then the saturated moment expected from the doublet ground state (1.56 $\mu_B$). Allowing for the Ru and Pr ordering being described by different IRs gave the same result. This suggests that down to 1.5 K there is no ordering on the Pr sublattice. The results of the fits to the 1.5 K data, listed in Table~\ref{tab:irrep_fits}, shown in Figure~\ref{irrep_fits} and discussed below, therefore only take into account an ordered moment on the Ru sublattice.
\begin{table}
	\caption{Corresponding BVs of the IRs given in Equation~\ref{eq:irreps}. The Ru/Pr atoms of the nonprimative basis are defined according to 1: (0, 0, 0)/(.5, .5, .5), 2: (.5, .75, .25)/(0, .25, .75), 3: (.25, .5, .75)/(.75, 0, .25) and 4: (.75, .25, .5)/(.25, .75, 0).}\label{tab:BV}
	\begin{tabular}{cc|cccccccccccc}\hline \hline
	IR & BV & \multicolumn{12}{c}{BV components} \\
	 & & \multicolumn{3}{c}{atom 1} & \multicolumn{3}{c}{atom 2} & \multicolumn{3}{c}{atom 3} & \multicolumn{3}{c}{atom 4}\\
	& & m$_a$ & m$_b$ & m$_c$ & m$_a$ & m$_b$ & m$_c$ & m$_a$ & m$_b$ & m$_c$ & m$_a$ & m$_b$ & m$_c$\\ \hline
	$\Gamma_3$ & $\psi_1$ & 1 & 1 & 1 & -1 & -1 & 1 & -1 & 1 & -1 & 1 & -1 & -1\\
	$\Gamma_6$ & $\psi_2$ & 2 & -1 & -1 &	-2 & 1 & -1 & -2 & -1 & 1 & 2 & 1 & 1\\
	 & $\psi_3$ & 0 & -1 & 1 & 0 & 1 & 1 & 0 & -1 & -1 & 0 & 1 & -1\\
	$\Gamma_8$ & $\psi_4$ & 1 & -1 & 0 & -1 & 1 & 0 & 1 & 1 & 0 & -1 & -1 & 0\\
	 & $\psi_5$ & 0 & 1 & -1 & 0 & 1 & 1 & 0 & -1 & -1 & 0 & -1 & 1\\
	 & $\psi_6$ & -1 & 0 & 1 & -1 & 0 & -1 & 1 & 0 & -1 & 1 & 0 & 1\\
	$\Gamma_{10}$ & $\psi_7$ & 1 & 1 & 0 & -1 & -1 & 0 & 1 & -1 & 0 & -1 & 1 & 0\\
	 & $\psi_8$ & 0 & 0 & 1 & 0 & 0 & 1 & 0 & 0 & 1 & 0 & 0 &1\\
	 & $\psi_9$ & 0 & 1 & 1 & 0 & 1 & -1 & 0 & -1 & 1 & 0 & -1 & -1\\
	 & $\psi_{10}$ & 1 & 0 & 0 & 1 & 0 & 0 & 1 & 0 & 0 & 1 & 0 & 0\\
	 & $\psi_{11}$ & 1 & 0 & 1 & 1 & 0 & -1 & -1 & 0 & -1 & -1 & 0 & 1\\
	 & $\psi_{12}$ & 0 & 1 & 0 & 0 & 1 & 0 & 0 & 1 & 0 & 0 & 1 & 0\\ \hline
	\end{tabular}
\end{table}

Of all four possible IRs only $\Gamma_8$ does not want to fit to the data. This can be explained by the fact that while for this model the calculated Bragg intensities of the (111) and (002) reflections are close to being equal in the data no magnetic intensity is observed for the (002) reflection (Fig.~\ref{irrep_fits}(c)). While $\Gamma_3$ and $\Gamma_{10}$ do fit to the data slightly better it can be seen from Figs.~\ref{irrep_fits}(a) and (d) that they only allow for magnetic intensity on the (220) and (111) reflection respectively (in the shown $2\theta$ range). As both reflections are observed, these IRs do not describe the observed ordering. $\Gamma_6$ has 2 associated BVs (with moments either off-diagonal (along [211]) or co-planar (along [011])), fitting each individually gives an identical fit to the data describing all the observed magnetic Bragg scattering (Fig.~\ref{irrep_fits}(b) shows the fit of $\psi_2$ to the data). Due to the powder averaging we are unable to distinguish between these two BVs and/or determine whether the actual magnetic structure is a combination of the two. For this a single crystal diffraction experiment will be needed. 
\begin{table}
	\caption{Refined magnetic parameters from fits of the individual IRs given in Equation~\ref{eq:irreps} to powder neutron diffraction profiles of the $x= 0$ sample collected at 1.5 K. For the fits of $\Gamma_8$ to the data the size of the ordered moment was fixed to 1.41 $\mu_B$. For $\Gamma_8$ and $\Gamma_{10}$ only $\psi_4$ and $\psi_7 + \psi_8$ respectively were fitted to the data as the other associated BVs are related by alternative choice of lattice axis.}\label{tab:irrep_fits}
	\begin{tabular}{c|ccccc}\hline \hline
	IR & $\Gamma_3$ & \multicolumn{2}{c}{$\Gamma_6$} & $\Gamma_8$ & $\Gamma_{10}$\\
	BV & $\psi_1$ & $\psi_2$ & $\psi_3$ & $\psi_4$ & $\psi_7 + \psi_8$\\ \hline
	R$_p$ (\%) & 11.1 & 10.7 & 10.7 & 11.2 & 11.0 \\
	R$_{wp}$ (\%) & 8.94 & 8.62 & 8.62 & 9.21 & 8.84 \\
	R$_{magn}$ (\%) & 6.82 & 5.52 & 5.49 & 49.6 & 18.5 \\
	$\chi^2$ & 2.048 & 1.905 & 1.905 & 2.169 & 2.002 \\
	$\mu_{\mathrm{Ru}}$ ($\mu_B$) & 1.38(5) & 1.49(5) & 1.47(3) & 1.41 & 1.50(11) \\ \hline
	\end{tabular}
\end{table}

\begin{figure}
	\includegraphics[width=8.0cm]{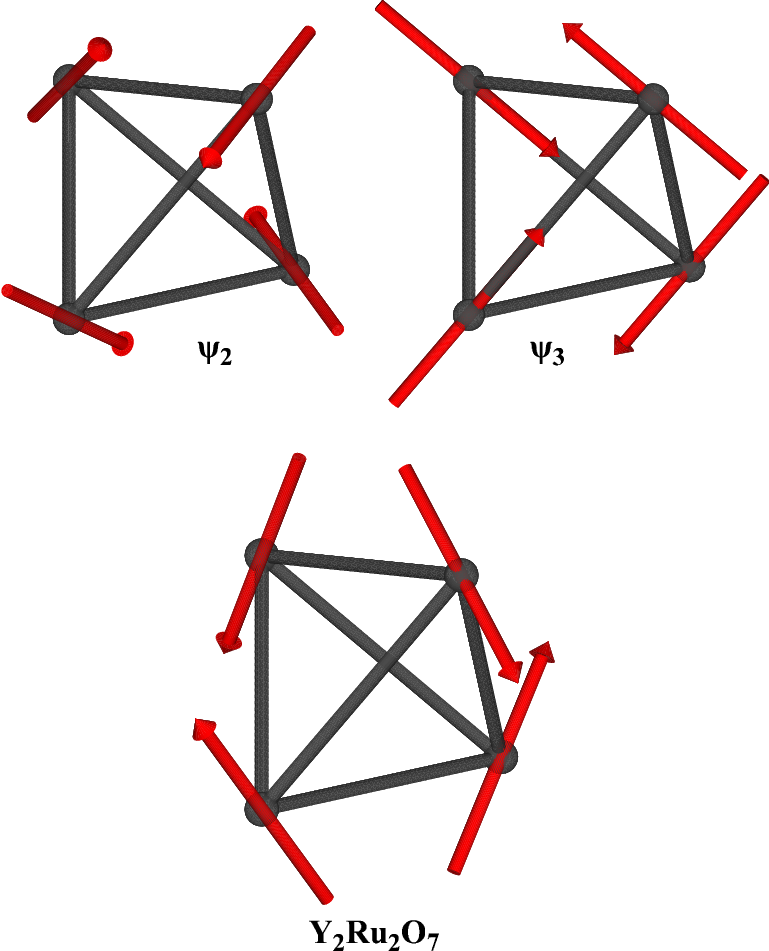}
	\caption{Alignment of the magnetic Ru moments, according to $\psi_2$, $\psi_3$ and Y$_2$Ru$_2$O$_7$, within a single single tetrahedra.~\cite{Y2Ru2O7,vesta}}
	\label{magn_struct}
\end{figure}

In summary our analysis has shown that describing the ordering of the Ru moments in $x= 0$ either by the model proposed for Y$_2$Ru$_2$O$_7$ or by the IR $\Gamma_6$ of space group  $Fd\bar{3}m$ gives identical fits to the data (Figs.~\ref{d_y2ru27_fit} and~\ref{irrep_fits}). Closer examination of both models reveals that the two associated BVs of $\Gamma_6$ are special cases of the more general description used for Y$_2$Ru$_2$O$_7$ and these models are therefore identical (Fig.~\ref{magn_struct}).~\cite{Y2Ru2O7} Unlike what is found for the other magnetic rare earth containing Ru-pyrochlores this suggests that the the ordering of the Ru moments in $x=0$ is not influenced by the Pr rare earth anisotropy.~\cite{Ho2Ru2O7,Er2Ru2O7} This in turn implies that the Pr ground-state is non-magnetic and can explain why no ordering associated with the Pr sublattice is observed down to 100 mK. Interestingly the ordering observed for $x=0$ (and Y$_2$Ru$_2$O$_7$) is the same as found for Er$_2$Ti$_2$O$_7$, there the rare earth anisotropy selects co-planar ordering ($\psi_3$).~\cite{Er2Ti2O7}

\subsubsection{Temperature dependence}
\begin{figure}
	\includegraphics[width=8.5cm]{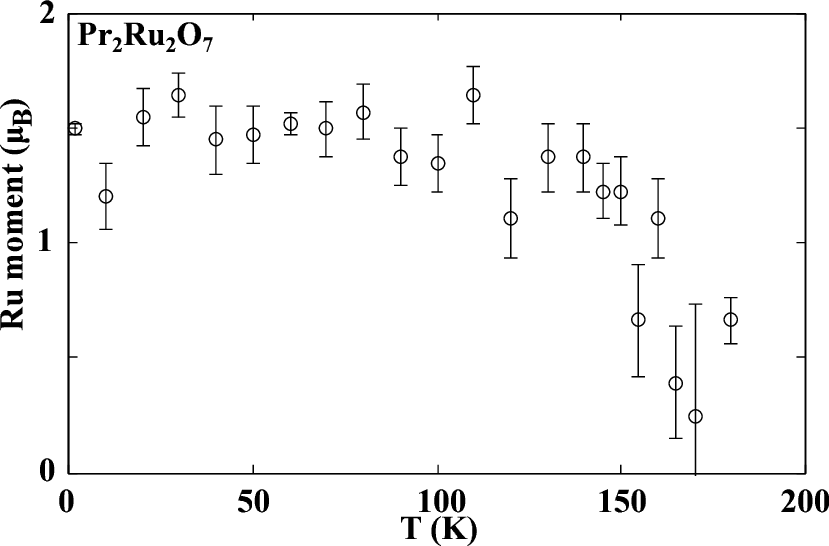}
	\caption{Temperature dependence of the ordered Ru moment.}
	\label{tdep_om}
\end{figure}

Figure~\ref{tdep_om} shows the temperature dependence of the ordered Ru moment, which was obtained by fitting the neutron power diffraction data using $\psi_2$ to describe the Ru ordering. This clearly shows the onset of an ordered Ru moment below 170 K confirming that the anomaly observed in both the specific heat and magnetization measurements at 165 K is associated with the ordering of the Ru-sublattice.~\cite{Pr2Ru2O7_DC,Pr2Ru2O7_Cp} As the temperature is further decreased the size of the ordered moment increases, to level off below 100 K to a value of ~1.5 $\mu_B$. This is in line with what is observed in the other Ru-pyrochlores.~\cite{Ho2Ru2O7,Y2Ru2O7,Er2Ru2O7} 

\subsection{Static disorder}

\subsection{Low energy magnetic response}
Comparing the low energy magnetic response of the pure ($x= 0$) with that of the previously measured Bi-doped material ($x= 0.97$), Figure.~\ref{iris_cont},  one observes the following. In both samples the wave vector dependence of the dispersion indicates that it is associated with Pr$^{3+}$ single ion physics and must arise from splitting of the doublet ground state. However, while for $x= 0$ the peak in the dispersion is rather sharp (but not resolution limited) and temperature dependent, for $x= 0.97$ it is broad and temperature independent. The presence of lone pairs on the Bi in $x= 0.97$ (non-magnetic and close to the insulator to metal transition) result in a low density of extended defects or a density wave which generates a continuum of local Pr environments that is temperature independent. 

The EXAFS results show that the environment about Bi is highly disordered for
both the $x=0.97$ and $x=2$ samples. This disorder on those A sites occupied by
Bi will modify the CF experienced by neighboring Pr ions. Since Pr is a
non-Kramers ion (and the determined doublet ground state relies on $D_{3d}$
point group symmetry) this distribution of local distortions will lead to a
broadened or split ground state. In the pure material this scenario with Bi is not
possible, but the EXAFS data indicate a small, but significant, disorder of the Pr atoms along
the Pr-O2 axis which may also spit the ground state.
Another possible scenario that might explain the
splitting of the Pr doublet ground state in the $x=0$ material is that it
results from the Ru-sublattice ordering.


It has already been observed for Y$_2$Ru$_2$O$_7$ that magnetoelastic effects play an important role in allowing the ordering of the Ru-sublattice to occur.~\cite{Y2Ru2O7_IR,Y2Ru2O7_PRB} While no optical data is available for $x= 0$ the ordering observed is identical to that of Y$_2$Ru$_2$O$_7$, as such it is very likely that the observed spin-phonon interactions that allow for the ordering in Y$_2$Ru$_2$O$_7$ are also present in $x= 0$. This combined with the lowering of the symmetry due to the Ru ordering can result in a distribution of internal fields (exchange and/or strain) which in turn leads to the observed splitting of the doubled ground state. 

\section{Conclusions} 
We have performed EXAFS, elastic and both high and low energy 
inelastic neutron scattering measurements on Pr$_{2-x}$Bi$_x$Ru$_2$O$_7$ to
understand the magnetic properties of this materials and complete those already
reported by us.~\cite{prbiru2o7_PRL} The EXAFS measurements reveal that the Ru
environment (B-site) remains well ordered throughout the series. In the case of
the A-site, the Pr environment has some small, but significant, intrinsic disorder
along the Pr-O2 axis which likely contributes to the splitting of the Kramers
doublet. In contrast, the environment about Bi is highly
disordered, and is attributed to the $6s$ lone pairs on Bi$^{3+}$, which
result in an off-center displacement. In agreement with previously reported
diffraction studies the Bi appears to move midway between two O1 atoms. Our CF
measurements reveal that the Pr ions have a doublet ground state and singlet
first excited state  ($\psi_g= |\mp4> - |\pm2> \pm |\mp1>$ and $\psi_1= |3> +
|0> + |-3>$ respectively). It was confirmed that this is also the case for
Pr$_2$Ir$_2$O$_7$. The high energy inelastic neutron scattering data also
suggest that strong CF-phonon coupling is present in both systems. Fits to the
diffraction data of the $x=0$ end member reveal no evidence (within our
experimental resolution) of a structural distortion associated with the
ordering of the Ru moments below T$_N$. The magnetic ordering of the Ru
sublattice is similar to that of Y$_2$Ru$_2$O$_7$ (or by IR $\Gamma_6$
of space group $Fd\overline{d}m$).~\cite{Y2Ru2O7} The ordering of the Ru
moments is not influenced by the Pr rare earth anisotropy and no ordering of
the Pr sublattice was observed down to 1.5 K. The low energy magnetic response
of Pr$_{2-x}$Bi$_x$Ru$_2$O$_7$  shows the presence of of a (broad) dispersion
associated with the splitting of the Pr doublet ground state. For $x=0$ it is
found to be temperature dependent, while for $x=0.97$ it is not. The nature of
the splitting of the (non-Kramers) doublet ground state changes upon doping,
going from being intrinsic and/or magnetoelastically induced in $x=0$ to result from Bi induced
A-site disorder in $x=0.97$. These measurements show that the Pr ground-state
can be very sensitive to local perturbations (be they ex- or intrinsic), something that needs to be taken
into account when studying materials containing this and other non-Kramers rare earth ions, e.g. Tb$2$Ti$_2$O$_7$, Hg$_2$Ti$_2$O$_7$ or LiHo$_x$Y$_{1-x}$F$_4$.~\cite{Tb2Ti2O7_2, Ho2Ti2O7, LiHoYF4}

\begin{acknowledgments}
The work at JHU and ISIS was supported by the US Department of Energy, office of Basic Energy Sciences, Division of Material Sciences and Engineering under  DE-FG02- 02ER45983 through 2008 and beyond that on DE-FG02-08ER46544. The work at Rutgers was supported by NSF through Grant No. DMR-0103858 and by the DOE under Grant No. DE-FG02-07ER46382. Work done at the UCLM was supported by the Ram\'{o}n y Cajal program through Grant no. RYC-2005-001064 and the Consejer\'{i}a de Educaci\'{o}n y Ciencia of the Junta de Comunidades de Castilla-La Mancha through Grant no. PII1I09-0083-2105.
K. H. K. was partially supported by KOSEF through CSCMR. The EXAFS experiments were performed at SSRL, operated by the DOE, Division of Chemical Sciences.
\end{acknowledgments}

\end{document}